Pre-publication version

*Assuring Autonomy International Programme*

# TIGARS

## Towards Identifying and closing Gaps in Assurance of autonomous Road vehicleS

Part 1

Overview and recommendations with a collection of Tigars Technical Notes on

Part 1

Assurance – overview and issues
Resilience and Safety Requirements
Open Systems Perspective
Formal Verification and Static Analysis of ML Systems

Part 2
Simulation and Dynamic Testing
Defence in Depth and Diversity
Security-Informed Safety Analysis
Standards and Guidelines

ADELARD

Pre-publication version

Editors & Author(s):
Robin Bloomfield, Adelard LLP and City, University of London
Gareth Fletcher, Adelard LLP
Heidy Khlaaf, Adelard LLP
Philippa Ryan, Adelard LLP

Contributing Author(s):
Shuji Kinoshita, Kanagawa University
Yoshiki Kinoshita, Kanagawa University
Makoto Takeyama, Kanagawa University
Yutaka Matsubara, Nagoya University
Peter Popov City, University of London
Kazuki Imai, Witz Corporation
Yoshinori Tsutake, Witz Corporation

# TIGARS

Towards Identifying and closing Gaps in Assurance of autonomous Road vehicleS

**Project Ref: 01/18/05**

**Adelard Ref: D/1259/138008/7**

24 Waterside
44–48 Wharf Road
London
N1 7UX

T +44 20 7832 5850
F +44 20 7832 5870
E office@adelard.com
W www.adelard.com

Authors
Robin Bloomfield
Heidy Khlaaf

# TIGARS OVERVIEW AND INTRODUCTION TO THE TIGARS TOPIC NOTES

## Summary

TIGARS aims to improve the assurance of the first-generation autonomous vehicles (AVs), or more generally Robotics and Autonomous Systems (RASs), currently being deployed, and how existing approaches for assurance need to change to address current and future autonomous systems. The project provides a cross-sector and international perspective. It is a partnership between Adelard LLP, City University in London, the University of Nagoya, Kanagawa University, and WITZ Corporation.

We argue that assuring trust and trustworthiness through argument-based mechanisms, specifically, the Claims, Arguments, and Evidence (CAE) framework, allows for the accelerated exploration of novel mechanisms that could lead to advancements in the assurance of disruptive technologies. This assurance approach is informed by an understanding of engineering processes and technical analysis for developing and assuring autonomous vehicles addressing: resilience, formal verification, static analysis, security, and other aspects. This report provides an introduction and overview of the TIGARS Topic Notes (TTNs) to support the development and evaluation of autonomous vehicles.

## Use of Document

## Acknowledgement

This project is partially supported by the Assuring Autonomy International Programme, a partnership between Lloyd's Register Foundation and the University of York. Adelard acknowledges the additional support of the UK Department for Transport.

## Contents

# 1 Introduction

TIGARS aims to improve the assurance of the first-generation autonomous vehicles (AVs), or more generally Robotics and Autonomous Systems (RASs), currently being deployed, and how existing approaches for assurance need to change to address current and future autonomous systems. The project provides a cross-sector and international perspective. It is a partnership between Adelard LLP, City University in London, the University of Nagoya, Kanagawa University, and WITZ Corporation.

We argue that assuring trust and trustworthiness through argument-based mechanisms, specifically, the Claims, Arguments, and Evidence (CAE) framework, allows for the accelerated exploration of novel mechanisms that could lead to advancements in the assurance of disruptive technologies. This assurance approach is informed by an understanding of engineering processes and technical analysis for developing and assuring autonomous vehicles addressing: resilience, formal verification, static analysis, security, and other aspects. Our project

- Identifies current autonomous systems engineering approaches and their assurance gaps. We assess the current state of software engineering development practice and the feasibility of deploying current state-of-the-art static analysis, verification, and testing techniques.
- Investigates how to address the assurance gaps with new analysis approaches based on verification of machine learning in both benign and adversarial environments, using simulation and test strategies, and an evaluation of defence in depth.
- Provides recommendations to regulatory and policy organisations and standards developers on a principles-based framework to address autonomy, as well as on a near-term interpretation of existing standards.

From this perspective, we have produced a number of TIGARS Topic Notes (TTNs) to support the development and evaluation of autonomous vehicles.

The TTNs address the challenges faced in the current landscape regarding the noted attributes for ML-based autonomous vehicles and systems. Additionally, we discuss potential solutions and recommendations proposed by a varied set of literature as well as preliminary research that we have carried out. The accompanying TTNs are

- Assurance – Overview and Issues [1]
- Resilience and Safety Requirements [2]
- Open Systems Perspective [3]
- Formal Verification and Static Analysis of ML Systems [4]
- Simulation and Dynamic Testing [5]
- Defence in Depth and Diversity [6]
- Security-Informed Safety Analysis [7]
- Standards and Guidelines [8]

The autonomous systems field is international and has a wide variety of players of differing maturity. Some entrants are unfamiliar with classical safety engineering, yet have expertise related to AI and ML-based systems. Others are mature and familiar with classical assurance approaches but lack a grasp on the challenges autonomy brings about. Given this wide range of maturity and backgrounds, the TIGARS outputs aim to address a range of different audiences. We hope they will be accessible and of interest to both engineers as well as policy makers. Below, we provide conclusive remarks and overall recommendations derived from each TTN.

This work is part of the Assuring Autonomy International Programme. In Appendix A, we relate each TTN to the Body of Knowledge being developed in that programme.

# 2 Conclusive remarks and recommendations

## 2.1 Assurance cases and approaches

1. Developing an assurance strategy should be a key part of the overall design approach and integrated into the overall lifecycle. The assurance approach should be commensurate with the different risks be consistent across them, e.g., by adopting an outcome-based risk informed approach.
   1.1. Novel assurance approaches (e.g., articulated using CAE) exclusive to ML and AI-based systems should be developed to identify areas to focus on and establish how they impact both the system and its assurance. It can help define and evaluate the reasoning and evidence needed.
   1.2. Key claims should address the high-level functional and ethical principles such as those from the EU Expert Group report [9] and the Sherpa project [10]. These principles can be used to shape and define system or service level properties.
   1.3. An assurance case for autonomous systems should at a minimum address the points below:
      - what the system is and in what environment and ecosystem it will operate in
      - how much trust in a system is needed, considering interdependencies and systemic risks
      - whether it is sufficiently trustworthy to be initially deployed
      - whether it will continue to be trustworthy in the face of environmental changes, threat evolution and failures
2. Structured argumentation for safety cases (and more generally assurance cases) needs more emphasis on reasoning and evidence, if the cases are to be sufficiently robust and acceptable. We have characterised a new CAE based assurance framework to achieve this, which would utilize evidence extracted from V&V, defence in depth, and diversity techniques.

## 2.2 Resilience and safety analysis

AVs deployed as part of an ecology of systems that deliver services (e.g., mobility service), must appropriately define resilience and safety requirements. To develop the dependability or resilience of a service, discussions of requirements and assurance should start from a service level, not from systems or components level. Discussions should include vehicle capabilities, infrastructure sensors, cloud systems, etc. Resilience requirements should also be derived at service level, and then assigned to each system or component. The system should have high-level safety, security and resilience requirements. A systems theoretic approach (e.g., using STPA) combined with an impact of variability using FRAM can be useful to addressing these requirements.

## 2.3 Open systems dependability perspective

1. For AVs to be accepted socially, stakeholders need to have confidence, before they are deployed, in how they are going to adapt to changes post-deployment. This is particularly important when considering security requirements, as AVs are deployed in threat-filled environments that keep changing. The future behaviours of AVs should be assured systematically through Open Systems Dependability (OSD) deployed on the system's lifecycle.
2. Policy makers and regulators should enable and promote adoption of standardised OSD by AV manufacturers, operators and users. They play a key role in ensuring such collaboration is possible across legitimate AVs. Sharing of information, including assurance, is a major concern. Appropriate policies can incentivise AV manufacturers and operators to participate in a transparent evaluation and assurance regime which in turn strengthens users' confidence in RASs' future behaviours.

## 2.4 Verification and Validation (V&V) techniques

It is crucial to strategize the use of V&V (and defence in depth and diversity techniques in Section 2.5) through the lens of an assurance approach, in particular, CAE, to identify the role of such methods and how they complement other approaches. We provide recommendations regarding their use below:

Formal Verification

1. ML-specific properties such as pointwise robustness not only fail to note real-world examples, but also how state-of-the-art verification techniques can be applied to real-time systems. With regard to the safety assurance of an autonomous system, pointwise robustness fails to support or provide evidence for system robustness as discussed in [4]. We thus recommend
   1.1. Creation of relevant safety specifications unique to ML algorithms, with corresponding mathematical frameworks. The noted specifications must contribute to the assurance of an AI system, specifically, the context of an assurance case (i.e., CAE). Some ML algorithms (e.g., vision) may be intrinsically unverifiable against the properties which are of interest to the safety of an autonomous vehicle, however, other properties can in principle be formulated for other types of ML systems (e.g., planning) in autonomous vehicles.
   1.2. Collaboration between ML and verification researchers resulting in deep learning systems that are more amenable to verification. Novel formal verification techniques are needed which can address the newly defined specifications.

Static Analysis

2. The ML lifecycle relies heavily on data processed in a complex chain of libraries and tools traditionally implemented, often in Python. It has been demonstrated that implementation in these systems may propagate and affect the accuracy and functionality of the ML algorithm itself. We have demonstrated that static analysis tools can be used to build confidence in supporting systems. However, the verification of existing ML software infrastructure may pose particular challenges. We thus recommend:
   2.1. Creation of novel formally-based static analysis techniques addressing Python, and more generally, dynamically typed languages, given that they are not currently available. Formal methods can have a strong role in ensuring provenance of training and data processing.
   2.2. Organisations should consider rewriting any deployed safety critical software in a verifiable language if the appropriate analysis tools for Python are unavailable.
3. Organisations must understand the extent to which existing integrity static analysis tools can contribute to the confidence of the development of ML algorithms. The complexities arising from choice of implementation language, e.g., issues from using C or C++, should be well understood.

Simulation

4. The roles of the different simulation variants should be specified and justified, and confidence in the simulation environment needs to be established. This may include confidence in the modelled behaviour of the tested system, as well as confidence in the software running the simulation.
5. Attempts should be made to make the tests as repeatable as possible, however, if this is not possible the impact in confidence on the test results must be considered.
6. Adjustments in system behaviour may be needed to accommodate the simulation environments and these will need to be justified so that test evidence can be used in the overall assurance case.

## 2.5 Diversity and defence in depth

1. The use of diversity to improve reliability and safety is a sound principle. In particular it should be used to achieve higher dependability of safety mechanisms. The stakeholders for a mobility service or AVs should undertake a review of defence in depth and define a diversity and defence in depth strategy balancing the advantages of diversity with possible increases in complexity and attack surface.
2. Diversity should be considered within the construction of a system's architecture to reduce the trust needed in a single ML component. Independence of failures should not be assumed and failure correlation should be considered based where possible on experimental data. An architectural approach which limits reliance on sub-components of the system that need to be highly trusted (e.g., ML algorithms) should be taken.
3. Safety monitor architectures should be considered to reduce the trust needed in ML components as they monitor both the state of the environment and the AV. Where feasible, they can be used to gain performance and safety benefits of deploying complex ML components, whilst mitigating the risks of using such technologies, as discussed in [1].

## 2.6 Security-informed safety

1. Security-informed safety should be addressed at all stages of the innovation cycle from conceptualisation, experimentation, and prototyping through to production. A security-informed hazard analysis should be undertaken during development. The hazard analysis should be reviewed periodically during operation or when a safety related component has been updated or if additional threat and vulnerability information has been identified.
2. The UK PAS 11281 can be used to systematically consider security through addressing: security policy, organization and culture; security-aware development process; maintaining effective defences; incident management; and safe and secure design, all contributing to a safe and secure world. The PAS can be applied progressively during the innovation lifecycle of an AV or RAS, and adapted to provide a project specific implementation.
3. Security-informed safety cases are still novel, and the experience of developing and integrating security issues into the safety analysis should be captured. In the industry as a whole, more training and expertise for SIS analysis is required, as many decisions rely on expert judgement. Although methodology that has been developed in other sectors can also be applied to AVs, AI and ML based technologies will provide novel security challenges that must additionally be addressed.

## 2.7 Standardisation and guidance

1. Duplication of standardisation work on similar topics should be reduced to a minimum. Efforts to prevent duplication have been ongoing in the international standardisation community, but we have observed that AV and RAS relevant topics often have duplicate standards, which may not be aligned. An example of possible duplication is in risk management as described in [8].
2. An authoritative and introductory guideline covering necessary knowledge for AVs and RASs should be developed for new entrants to this arena. Particularly, guidelines should include surveys on foundational standards of the field. Many IT companies are entering into the market without the experience of the traditional manufacturers. The current lack of such overall guidelines can lead to IT stakeholders to overly concentrate on their strength within a particular area without essential knowledge of AI/ML or safety. The recommended guideline would help ensure that innovative technologies and traditional engineering and assurance practices are aligned.

# 3 Bibliography

[1] TIGARS Topic Note, Assurance – Overview and Issues, D5.6.1 (W3013). December 2019.

[2] TIGARS Topic Note, Resilience and Safety Requirements, D5.6.2 (W3035). December 2019.

[3] TIGARS Topic Note, Open Systems Perspective, D5.6.3 (W3036). December 2019.

[4] TIGARS Topic Note, Formal Verification and Static Analysis of ML Systems, D5.6.4 (W3014). December 2019.

[5] TIGARS Topic Note, Simulation and Dynamic Testing, D5.6.5 (W3015). December 2019.

[6] TIGARS Topic Note, Defence in Depth and Diversity, D5.6.6 (W3021). December 2019.

[7] TIGARS Topic Note, Security-Informed Safety Analysis, D5.6.7 (W3022). December 2019.

[8] TIGARS Topic Note, Standards and Guidelines, D5.6.8 (W3025). December 2019.

[9] High-Level Expert Group on Artificial Intelligence, Ethics guidelines for trustworthy AI, available online (https://ec.europa.eu/digital- single-market/en/high-level-expert-group-artificial-intelligence).

[10] Shaping the ethical dimensions of smart information systems – a European perspective (SHERPA), Deliverable No. 3.2. BSI ART/1_19_0257

# Appendix A
## Relationship to AAIP BoK

The following table maps the TTNs to the York BoK Sections and subsections:

| TIGARS Topic Note | York AAIP BoK |
|---|---|
| Assurance framework | This note does not map that readily to the BoK but the areas it does address are:<br>Section 4 on gaining approval including<br>4.2 Risk acceptance<br>4.3 Provision of sufficient confidence in required behaviour<br>3.2.2 Managing assurance deficits |
| Resilience and safety requirements | The BoK does not address resilience as a distinct topic. Relevant sections are:<br>Section 1 on defining required behaviour including<br>1.1 Identifying hazards<br>1.2 Identifying hazardous system behaviour<br>1.3 Defining safety requirements<br>1.4 Impact of security on safety – implicitly via resilience |
| Open Systems dependability perspective | OSD addresses the whole lifecycle, relevant Sections are 1, 2, 3, 4, and in particular, Section 2.6 Handling change during operation |
| Static analysis and formal verification of ML systems | 2.3 Implementing the requirements using ML<br>3.1.1 Identifying Sensing deviations<br>3.1.6 Identifying ML deviations |
| Testing and simulation | 2.3.3 Verification of the learned model<br>2.7 Using Simulation<br>3.1 Identifying potential deviation from required behaviour (test room)<br>3.1.6 Identifying ML deviations (ViViD)<br>3.2.1 Managing failures of machine-learnt components<br>3.2.2 Managing assurance deficits |
| Defence in depth and diversity | 2.2 Implementing of SUDA elements<br>3.2.1 Failure mitigation<br>3.1 Identification of potential deviation from required behaviour<br>3.1.6 ML deviations<br>4.3 Provision of sufficient confidence in required behaviour |
| Security informed safety | 1.4 Impact of security on safety<br>1.1 Identifying hazards<br>1.2 Identifying hazardous system behaviour<br>3.1 Identification of potential deviation from required behaviour<br>3.2.1 Failure mitigation |

| TIGARS Topic Note | York AAIP BoK |
|---|---|
| Standards and guidelines | 4.1 Conforming to rules and regulations<br>4.1.1 Identifying rules and regulations<br>4.1.2 Understanding requirements of rules and regulations |

# TIGARS

Towards Identifying and closing Gaps in Assurance of autonomous Road vehicleS

**Project Ref: 01/18/05**

**Adelard Ref: W3013/138008/19**

# TIGARS TOPIC NOTE 1: ASSURANCE - OVERVIEW AND ISSUES

## Summary

This paper discusses the assurance of autonomous vehicles through the lens of a structured assurance case. It describes an assurance case for a generic autonomous vehicle showing a thread from the top-level claims to the evidence of AI/ML based sensors. It also provides an introduction and context to the more specific public domain "TIGARS Topic Notes" (TTN) that we have produced. These cover a variety of topics: resilience and safety, security, learning and adaptation, defence in depth and diversity, and verification & validation addressing the assurance of machine learning algorithms, as well as a snapshot of the standards landscape.

24 Waterside
44–48 Wharf Road
London
N1 7UX

T +44 20 7832 5850
F +44 20 7832 5853
E office@adelard.com
W www.adelard.com

Authors
Robin Bloomfield
Heidy Khlaaf
Philippa Ryan
Gareth Fletcher

## Use of Document

The document is made available as a resource for the community, providing all use is adequately acknowledge and cited.

This document provides a snapshot of work in progress. We welcome feedback and interest in this work. Please contact the authors or admin.tigars@adelard.com

### Acknowledgement

This project is partially supported by the Assuring Autonomy International Programme, a partnership between Lloyd's Register Foundation and the University of York. Adelard acknowledges the additional support of the UK Department for Transport.

## Contents

## Figures

## Tables

# 1 Introduction and background

This paper discusses the assurance of autonomous vehicles through the lens of a structured assurance case. It aims to

- Explain how an assurance case could be developed and how existing approaches can be deployed to structure, reason, and integrate evidence into a case.
- Introduce the evolution and emphasis in our approach to assurance cases, and how they address the challenge of assuring AI/ML based components for use.
- Provide an introduction and context to the TIGARS Topic Notes (TTN), which aim to address the assurance gaps of new analysis approaches based on resilience and requirements, security, learning and adaptation. Furthermore, we address verification and validation techniques of machine learning in both benign and adversarial environments, using simulation and test strategies, and an evaluation of defence in depth.

In the safety area, safety cases are a well-known approach for describing whether a system is safe, how it might be hazardous and why that judgement can be trusted. When we are dealing with systems whose failure can lead to danger, a safety case is the appropriate approach. For subsystems and other services with only an indirect impact on safety, or for components of a safety relevant system, we need to have confidence that they meet their explicit or implicit requirements in a way that leads to the safety of the overall system. One approach to addressing the need for confidence in engineering systems and subsystems is to generalise the notion of safety case to an assurance case that provides justified confidence in the properties of interest (e.g., functionality, security, reliability, resilience).

The assurance of software-based components in the automotive sector often takes on a standards-based approach, but in the case of ML/AI based components, it is not possible to rely on standards. It is also questionable at a system level given the lack of validated standards, policies, and guidance for such novel technologies and their use. Nevertheless, standards are important in defining and promulgating good practice and shared terminology and concepts. An associated TTN reviewing and tracking standards is presented in [36]. Most system level standards and frameworks require some form of safety case or safety justification – see for example "Safety first", UL4600, ISO26262, UK PAS 12281, UK PAS. We focus on directly investigating whether the desired behaviour (e.g., safety property, resilience or reliability) of a system has been achieved.

A framework that can help us towards such a task is a *claim-based* approach. The key advantage of a claim-based approach is that there is considerable flexibility in how the claims are demonstrated, since different types of arguments and evidence can be used as appropriate. Such a flexible approach is necessary when identifying gaps and challenges in uncharted territory, such as the assurance of novel autonomous systems. In this paper, we use Claims, Argument, Evidence (CAE) (see Section 1.1) to develop an outline of an overall assurance case, proceeding from top-level claims, concerning an experimental autonomous vehicle and its social context, down to claims regarding the evaluation of subsystems, such as the ML model. The lens of the assurance case is used to identify gaps and challenges regarding techniques and evidence aimed at justifying desired system behaviours. These challenges are then further elaborated in specific TTNs.

CAE allows us to systematically describe the issues to be addressed, and to illustrate a thread of reasoning from claims to evidence. In Section 2, we describe challenges that autonomy brings about, rooted in the complexity of the case and ML-based issues. However, there remain significant new challenges that are associated with reasoning and evidence rather than structure. While the latter can be addressed by the use of CAE Blocks (see Section 1.1), there is a need to evolve and develop new assurance methodologies. We thus propose an evolved approach known as Assurance2 in Section 3, which despite maintaining the structure of typically used CAE Blocks, is strengthened with an explicit focus on the evidence and the reasoning in cases.

## 1.1 Background

Over the past decade there has been a move to develop an explicit claim- or goal-based approach to engineering justification and considerable work has been done on the structuring of engineering arguments (e.g. [2], [3] and [4]) and supporting standards (e.g. ISO/IEC 15026-2:2011 and [5]). Current safety case practice makes use of a basic approach that can be related to ideas originally developed by Toulmin [6] – claims are supported by evidence and an argument ("warrant") that links the evidence to the claim. There are variants of this basic approach that present the claim structure graphically such as goal structuring notation (GSN) [2] or CAE [3]. These notations [2] can be supported by tools [7] [8] that can help to create and modify the claim structure and also assist in the tracking of evidence status, propagation of changes through the case, and handling of automatic links to other requirements and management tools. A rigorous analysis of assurance cases is provided in [9].

The key elements of the Claims, Argument, Evidence (CAE) approach are

- *Claims*, which are assertions about a property of the system or some subsystem. Claims that are asserted as true without justification become assumptions and claims supporting an argument are called sub-claims.
- *Arguments* link the evidence of the claim; the reasoning rules need to justify the claim from the evidence.
- *Evidence* that is factual and used as the basis of the justification of the claim.

In order to support the use of CAE, a graphical notation is used to describe the interrelationship of the claims, arguments and evidence. In practice, the desired top claims we wish to make such as "the system is adequately safe" are too vague or are not directly supported or refuted by evidence. It is therefore necessary to develop them into sub-claims until the final nodes of the assessment can be directly supported (or refuted) with evidence.

The basic concepts of CAE are supported by the ISO/IEC 15026-2:2011 international standard and industry guidance [3]. The framework additionally consists of CAE Blocks that provide a restrictive set of common argument fragments and a mechanism for separating inductive and deductive aspects of the argumentation. These were identified by empirical analysis of actual safety cases [10]. The Blocks are

- Decomposition: Partition some aspect of the claim, or divide and conquer.
- Substitution: Refine a claim about an object into a claim about an equivalent object.
- Evidence incorporation: Evidence supports the claim, with emphasis on direct support.
- Concretion: Some aspect of the claim is given a more precise definition.
- Calculation or proof: Some value of the claim can be computed or proved.

Figure 1 illustrates CAE Block use:

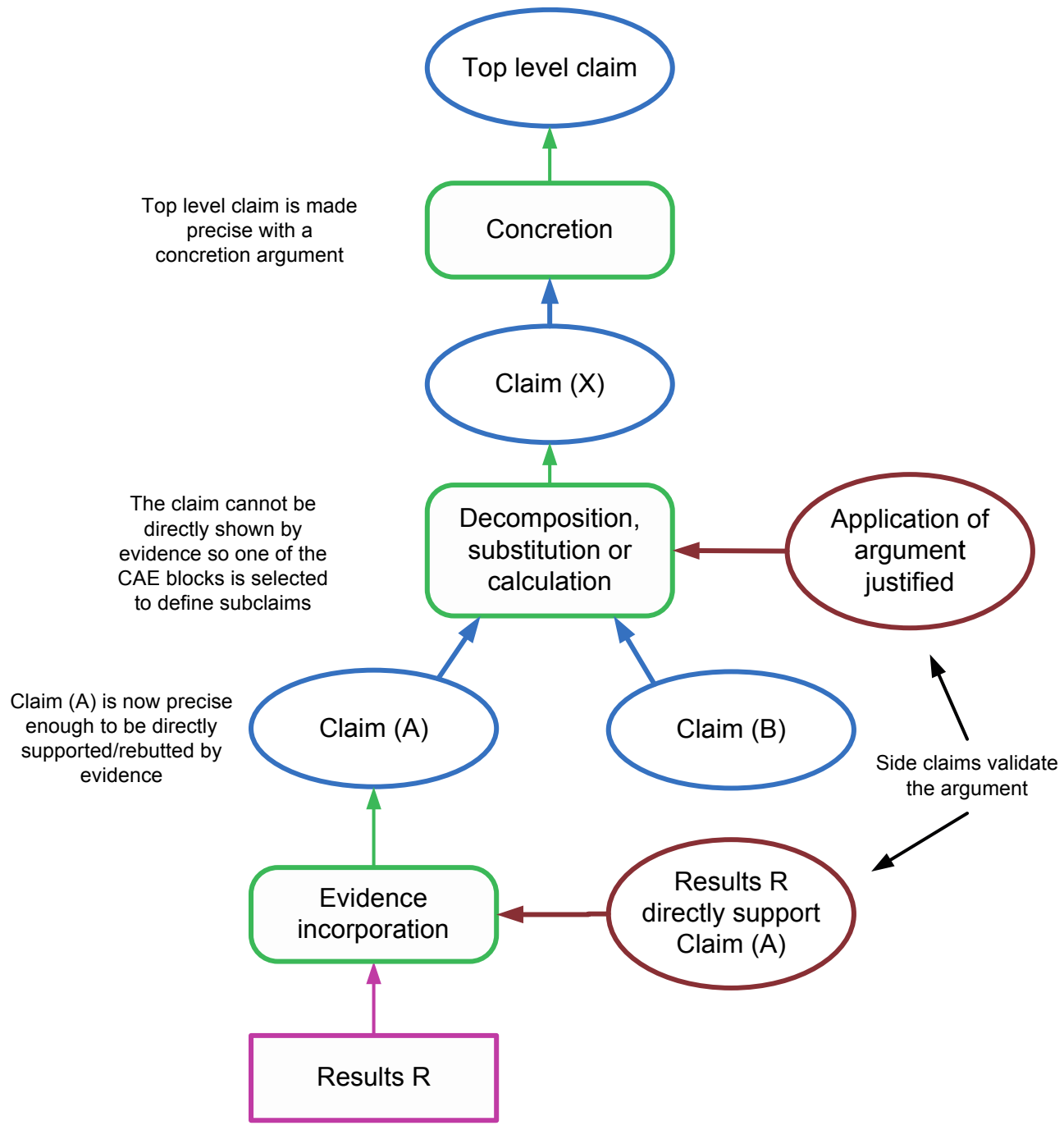

Figure 1: An example of CAE block use

One important aspect of the CAE Blocks is the use of side claims to justify how the sub-claims or evidence support the Block's top claim. In outlining the assurance case, we often omit the side claim for conciseness: in a real application they would need detailing and justifying.

The framework also defines connection rules to restrict the topology of CAE graphical structures. The use of Blocks and associated narrative can capture challenge, doubts and rebuttal and to illustrate how confidence can be considered as an integral part of the justification.

Technical background to the CAE framework and guidance material is available on the *https://claimsargumentsevidence.org* website.

## 2 Overall assurance perspective

Constructing an assurance case for an autonomous vehicle is a very complex task that as of now remains an open problem which many are pursuing [26][27]. This is due to the fact that a case would need to address a range of high-level properties (e.g., legal, safe, ethical, trustworthy, fair) in which the consensus of the properties' interpretations in the context of AI and ML is yet to be established and agreed on by multiple communities. Furthermore, one would need to provide a range of evidence including testing, formal verification, and simulations unique for ML, that may have not yet been developed or are currently not possible. Nonetheless, the key advantage of a claim-based approach is that there is considerable flexibility in how the claims are demonstrated, since different types of arguments and available evidence can be used as appropriate.

Generally speaking, to assure a system one must identify

1. How much trust in a system is needed – further addressed in Section 2.1
2. Whether it is sufficiently trustworthy initially – addressed in Section 2.2
3. Whether it will continue to be trustworthy – discussed in Section 2.3

Each of the aforementioned points must be considered in part of a larger lifecycle of system development. In TIGARS, we use an instantiation of the open systems dependability (OSD) model, as it provides requirements and guidelines for system lifecycles of open systems to achieve dependability, as shown in Figure 2. OSD identifies and addresses four issues by means of four process views: consensus building, accountability achievement, failure response and change accommodation. The application of OSD is described in a supporting TTN [31] discussing appropriate lifecycle process required for development, failure response, and system changes and adaptations.

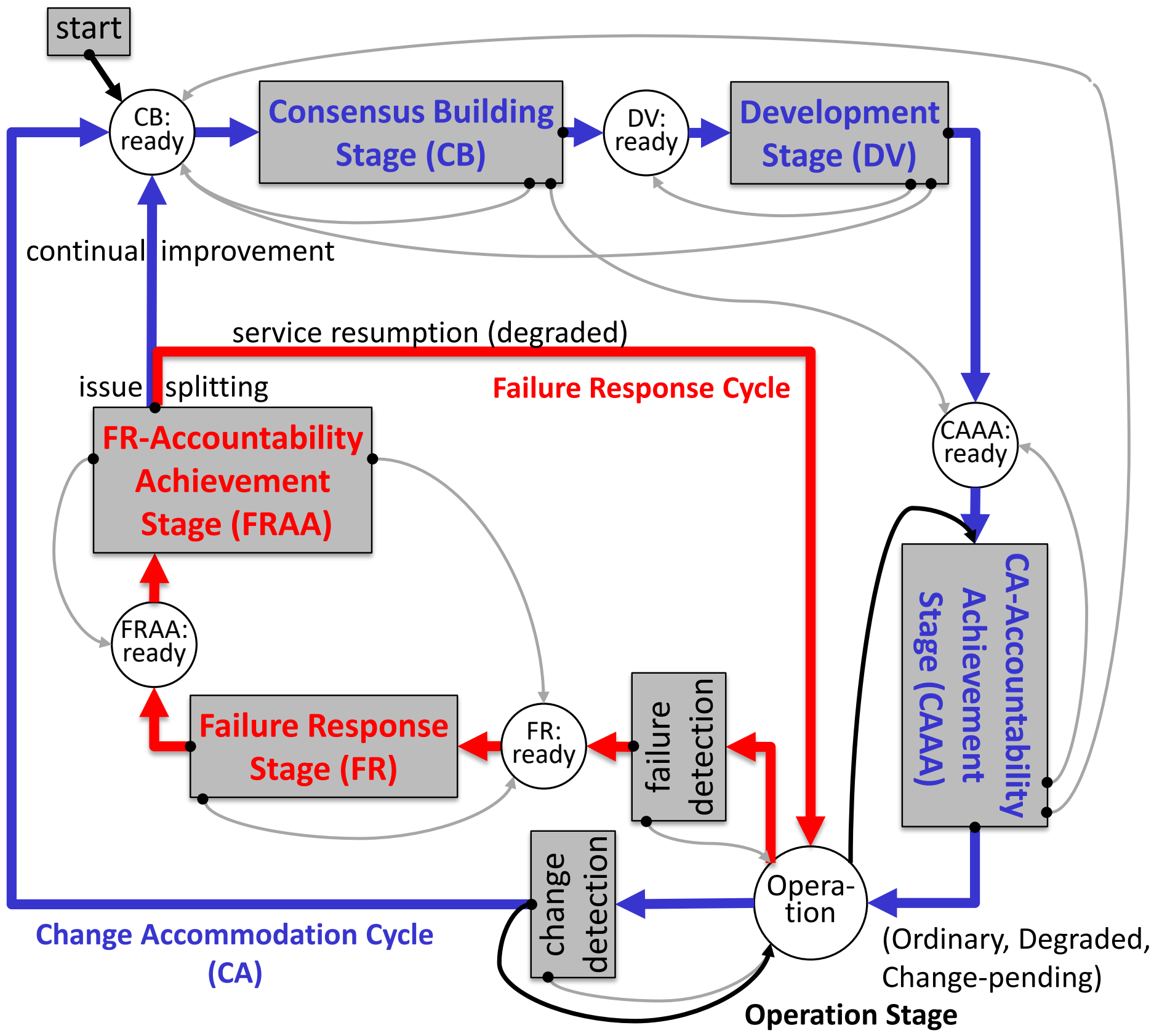

Figure 2: OSD lifecycle model

Considering an OSD approach, each of the above points must also respectively identify

1. That the organisation has sufficient competency, understanding, commitment and resources to deliver a trustworthy system. These will not be normative or deterministic, but their absence would give grounds for concern that need to be addressed on a specific project. A supply chain claim is also necessary to establish whether trust in the supply chain has been established. This is a key topic with many generic issues that is out of scope of the TIGARs project. However, governance requirements are detailed in PAS 11281 and include the attributes noted in Table 1 below:

| | |
|---|---|
| Policies and processes | Supply chain and other external dependencies |
| Responsibility and accountability | Security awareness and competency |
| Risk management | Culture and communication |
| Asset management | Protection of information |

Table 1: Topics to address in assessing policy, organisation, culture aspects

4. That the assurance case can sufficiently demonstrate that the evaluated autonomous vehicle or system can be deployed, i.e., a deployed system is sufficiently trustworthy initially. This would need substantial further justification in an actual case, as there will be changes to design, software versions, and the like, but is outside the scope of the TIGARS project.
5. That the deployed system will continue to meet its key requirements to maintain trustworthiness. In order for the top-level claims to be satisfied in the future, the system must be adaptable to changes, as defined by the OSD TTN [31].

Although the above points are well studied for traditional safety-critical systems (e.g., defence, nuclear, medical, etc.), they require reinterpretation or analysis for autonomous vehicles. We have thus developed a particular set of CAE structures that are generically applicable, and help identify how to develop trustworthy systems by explicitly considering evidence of sources of doubt, vulnerabilities, and mitigations addressing the behaviour of the system.

An overview of the case structure is shown in Figure 3. It outlines a structure from top-level claims regarding an autonomous vehicle and its environment (i.e., trust needed), down to sub-claims evaluating the safety and appropriateness of the AI/ML based subsystems (i.e., trustworthiness). The case additionally outlines claims required for a system to continuously meet its requirements in the future. There are a notable number of themes demonstrated within the case, ranging from requirements gathering, down to V&V techniques for ML-subcomponents. In the following sections, we examine each of the areas identified in Figure 3.

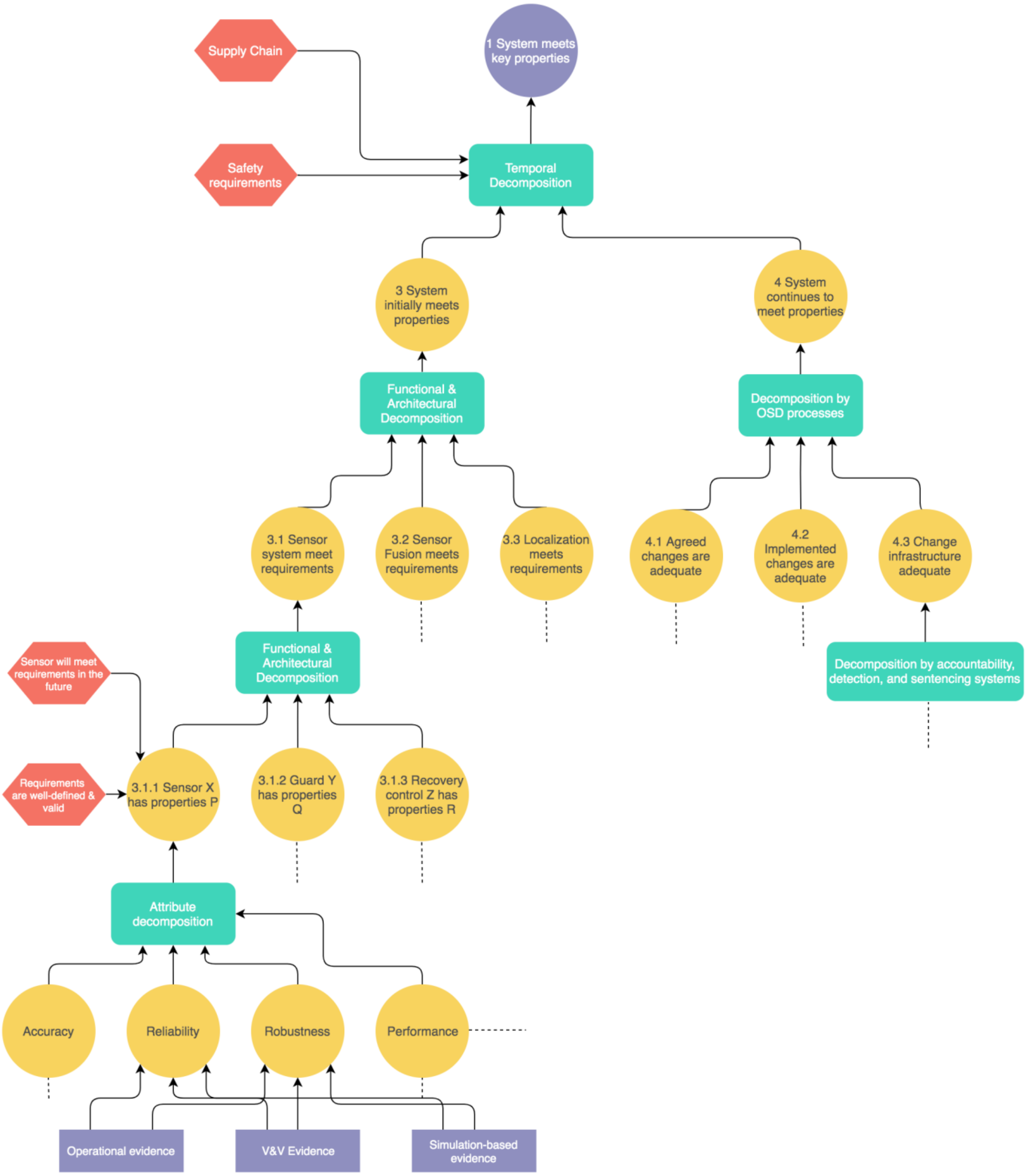

**Figure 3: Overview of assurance case template**

## 2.1 Requirements and resilience

Typically, the top-level claim of an assurance case addresses the trust needed in the system, this is demonstrated by claim 1 stating that there is high confidence that the "system meets key properties". Such a claim is vague and requires further details: what the system is – a vehicle or mobility service for example, and what do we mean by the key properties. For example, the key properties could address the high-level ethical principles in the EU Expert Group report [24] and those from the Sherpa project [25]. There are also industry specific principles [23]. These principles can be used to shape and define system or service level properties.

Depending on the use case for the vehicle there will be very different functional requirements and levels of performance required. Consider for example, a low speed automated vehicle in a restricted environment versus a cruise or chauffeur for motorway driving. Each should meet the relevant overarching ethical principles [24] and there should be consistency between the rigour required between the different applications as standards emerge for these different areas. Each vehicle will also differ in their minimum risk condition (MRC) [23], which defines when the vehicle must transition to manual use. Other aspects of resilience requirements such as availability and recovery need further exploration and development. The TIGARS project specifically focuses on STPA and FRAM techniques to support the definition of resilience requirements are described in the Resilience and Safety Requirements TTN [30].

Security is another important aspect of resilience and safety and will drive the need to consider systemic risks (e.g., of city-wide loss of mobility services), the need for adaptation and learning as well as requirements of rigour and governance. Each claim decomposition needs to consider security aspects: these issues are considered in the TTN on Security-Informed Safety [35].

In the next section, we consider how the resilience and safety requirements are initially met when the system is deployed, and in Section 2.3, how we can gain confidence (or not) that the requirements will be continue to be met.

## 2.2 System trustworthiness

Autonomous vehicles often contain a heterogeneous mixture of commercial-off-the-shelf (COTS) components including image-recognition, LIDAR, etc. Apportioning the trustworthiness, dependability, and requirements of each of these components in order to consider the real-time and safety related nature of the system is challenging. That is, we seek to develop and assure that the behaviour and performance of an ML component is sufficient to be effective. In Figure 3, claim 3 and below, we have constructed argumentation blocks within CAE to determine how architectures and sub-components allow for

1. emergent behaviour of the sub-components to correspond to the top-level claims, including that
   5.1. the required behaviour and functionality of the component are defined and valid
   5.2. the component behaves according to its requirements when deployed
   5.3. the component will carry on behaving according to its requirements for a future time frame
6. evidence (e.g., V&V) which contributes to the trustworthiness of component-specific claims
7. diversity and defence in depth to reduce the trust needed for specific ML components (discussed further in Section 2.2.2)

### 2.2.1 Systems architecture and ML subsystems

It is thus crucial to include an architectural decomposition analysis that defines component-specific claims for a particular subsystem. In this case, a CAE decomposition block is suitable to highlight the reasoning that allows us to compose component properties to justify the emergent behaviour of the overall system, such as that in Figure 4.

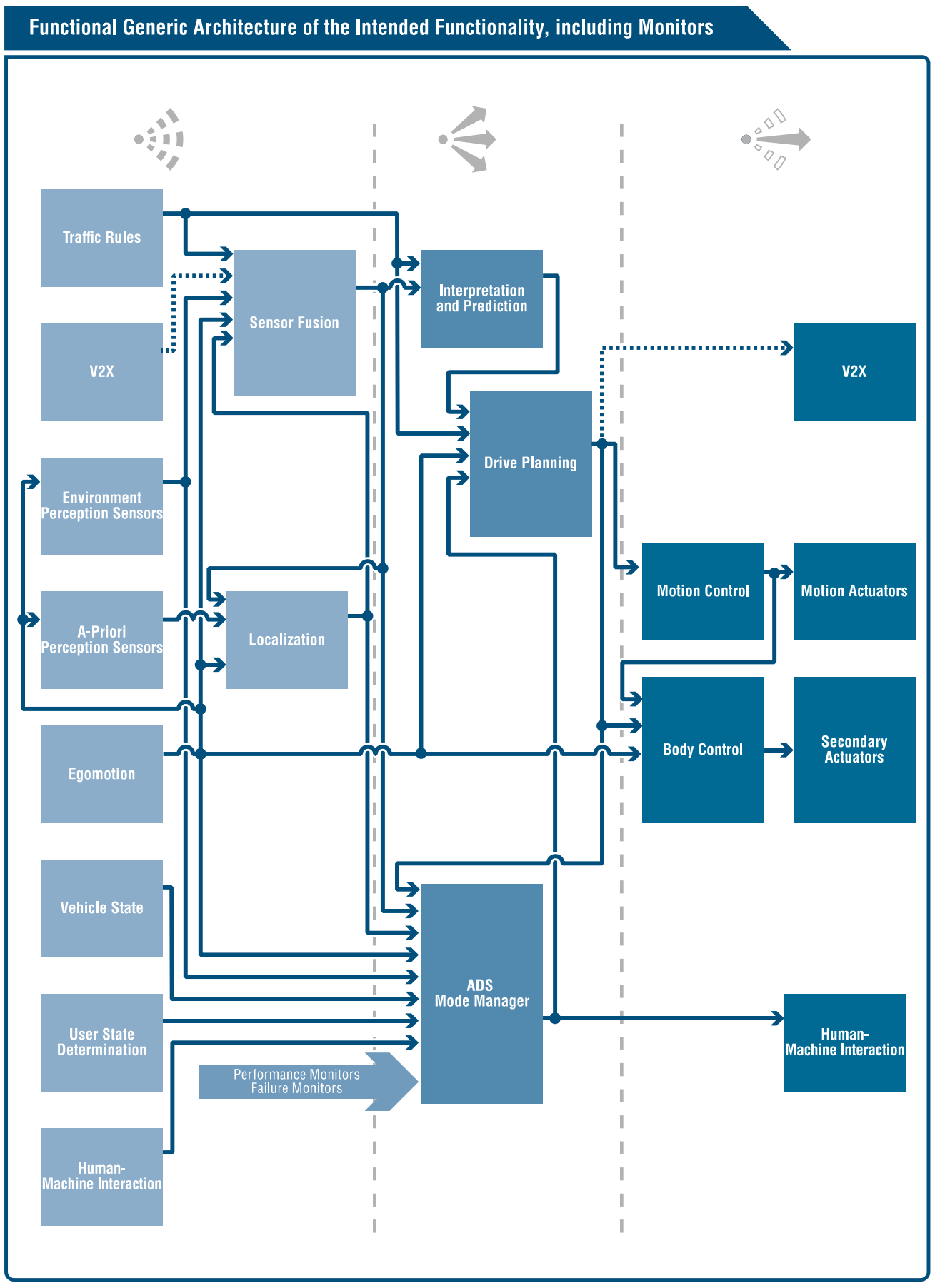

Figure 27: Functional Generic Architecture of the Intended Functionality, including Monitors

## Figure 4: Example overall architecture of an autonomous vehicle [23].

Initially, one may consider a decomposing the top-level claim in terms of different dependability attributes (e.g. timing, reliability) However, some properties may not be relevant at the component level (e.g., safety is a system property). Furthermore, not all subsystem claims will come from the refinement and apportionment of high-level requirements, but also from the requirement to support other parts of the case (e.g., supply chain assurance, future behaviour). In our case, we thus envisage a split into the platform and algorithm as reflected in Figure 5 and Figure 6 (corresponding to claims 3.1, 3.2, and 3.3 in Figure 3), in which an attribute split is applied *after* an architectural split.

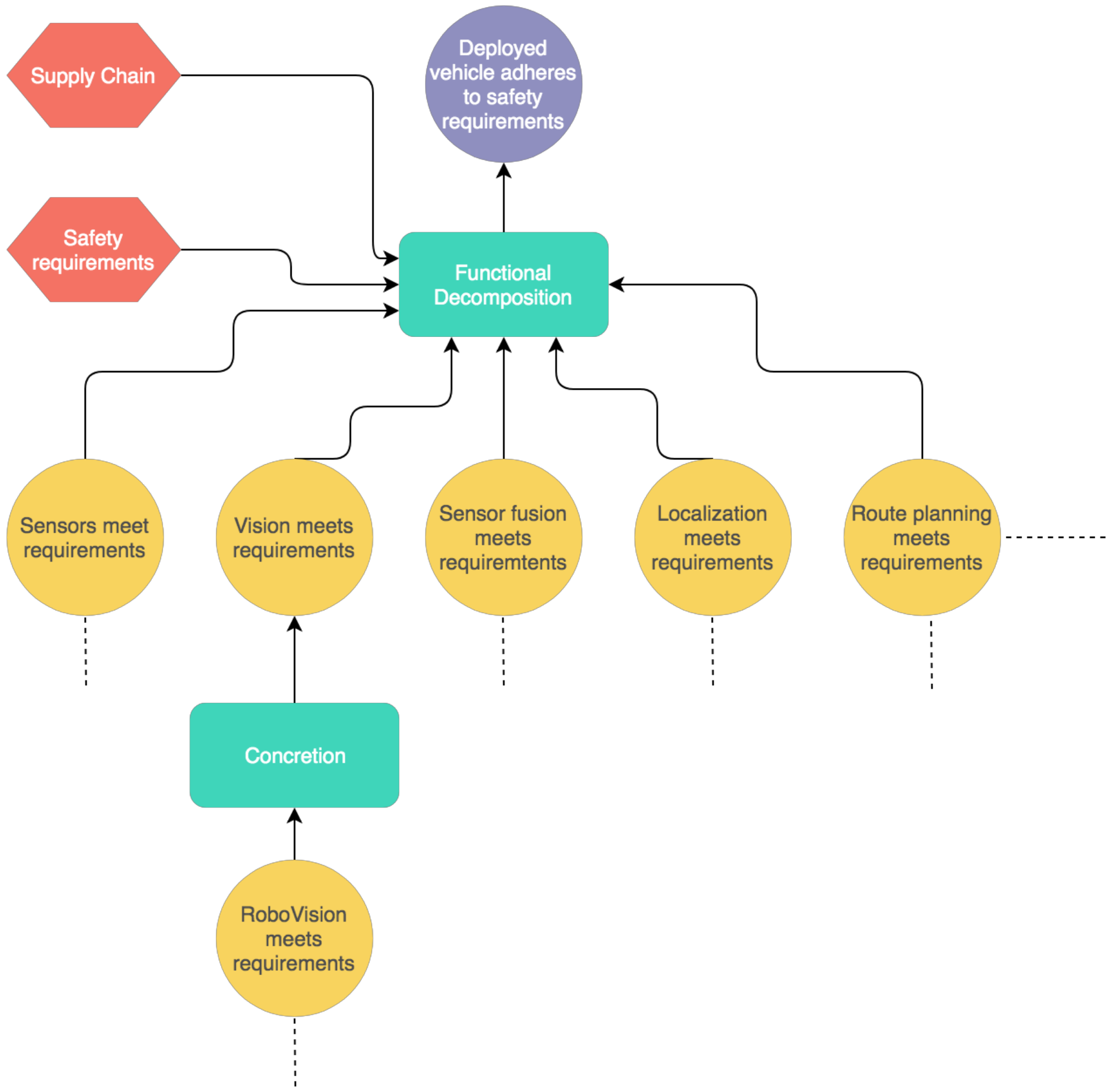

**Figure 5: A high-level example of an assurance sub-case in CAE regarding ML sub-components**

Whether an attribute or an architectural split is initially followed, we need to be able to trust the evidence that is produced, corresponding to the supporting claims of 1.1, 1.2, and 1.3 discussed in Section 2.2. We thus use the lens of the assurance case to identify gaps and challenges regarding techniques and evidence aimed at justifying desired system behaviours in the Formal Verification and Static Analysis of ML Systems TTN [32] and the Simulation and Dynamic Testing TTN [33]. In these notes, we identify that state-of-the-art V&V techniques for ML are unfortunately not mature enough to support behavioural claims for ML sub-components. We provide further recommendations and conclusions in each of the TTNs.

We note that other analyses are carried out at the architectural stage. A security risk analysis, exploring how a middle out architecture hazard analysis can investigate threats and identify mitigations and controls, is discussed in the TTN on Security-Informed Safety [35].

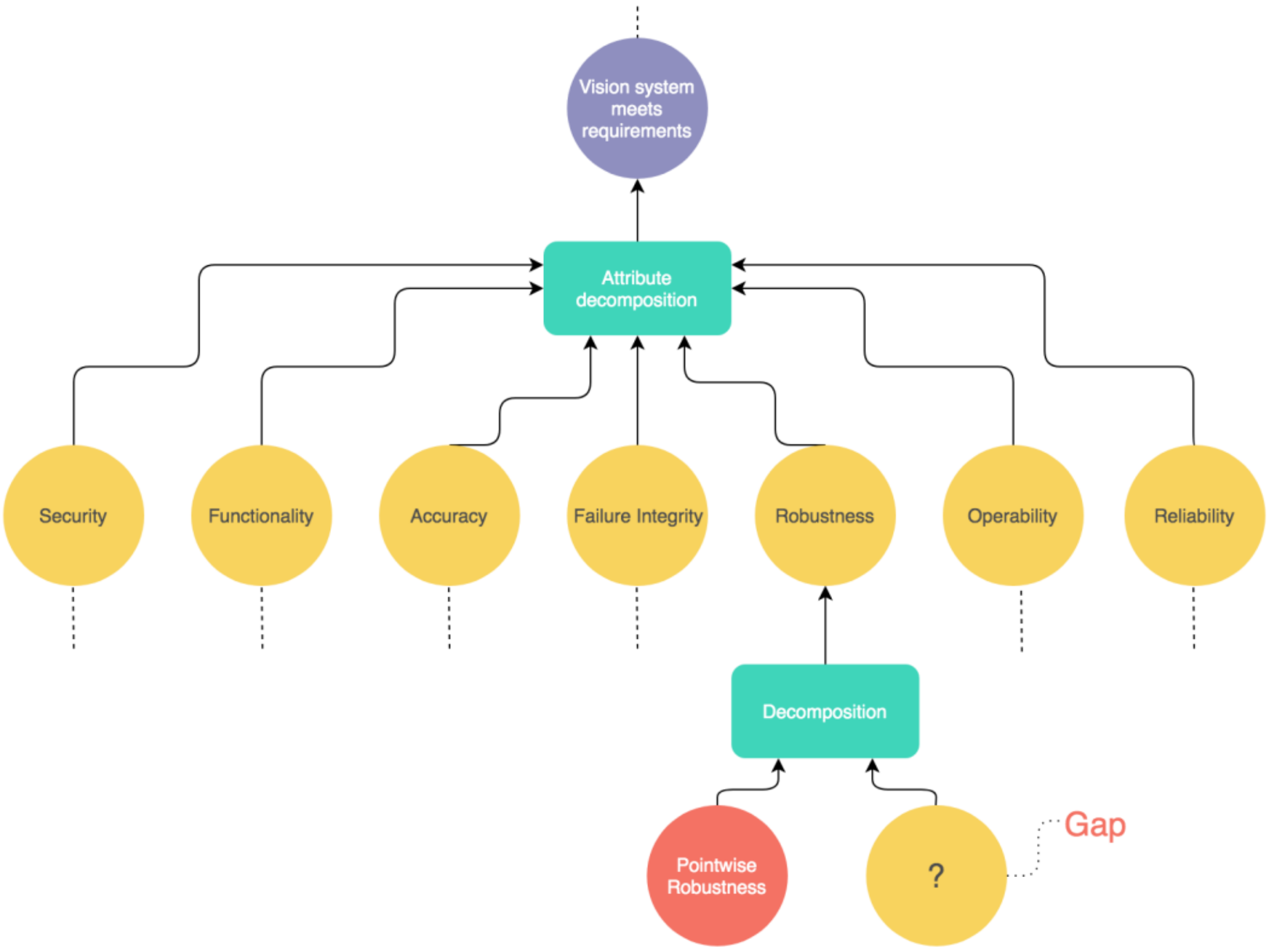

Figure 6: A sub-case demonstrating how CAE was used to identify component behaviour

### 2.2.2 Safety monitors and AI/ML component

Given that ML based components, especially perception systems, are difficult to assure, approaches are needed to reduce the assurance burden and allow their use. Consider engineered complex system architectures, which are used to limit parts of the system that need to be highly trusted: safety and security protection is provided in a simpler system or safety monitor that detects when a system is close to being in an unsafe or insecure condition, and acts accordingly. A safety monitor architecture is common across different disciplines (e.g., aircraft, railway systems, nuclear power plants, etc.) and is proposed as a standardised approach in the air domain [19], and more generally, for cyber physical systems [20]. In the remaining section, we address the impact of the use of a safety monitor within an assurance case.

Safety monitors can vary in sophistication from comparison between diverse sensors (e.g., comparison of LIDAR measured distance with that from a stereo camera) to a monitor implementing a complex set of equations and constraints (e.g., see Responsibility-Sensitive Safety (RSS) [21]). This architectural approach often seeks to reduce the trust needed in ML components by monitoring both the state of the environment and the vehicle. Their intention is to also monitor when an autonomous system is under stress, or in an error prone situation. It is not unlike the intrusion detection problem in security, where one tries to infer potentially dangerous behaviour from the complex system state and knowledge of threats. The DARPA Assured Autonomy programme for example, extends the safety monitor concept to include a dynamic assurance case, as monitors can be seen as form of run-time certification that shifts the certification or assurance challenge from design and development part of the lifecycle to operation [22].

We are thus particularly interested in how safety monitors can be used to gain the performance and safety benefits of deploying complex ML components, whilst mitigating the risks of using such technologies. Our approach aims to deploy an architecture that limits reliance on sub-components of the system that need to be highly trusted (e.g., ML algorithms). Instead safety and security protections are provided in a simpler system or safety monitor. In Figure 7 we have adapted the safety monitor architecture of [19] to include both a safety monitor and a complex function monitor, implemented for an AI/ML based system (note that we use the term AI/ML rather than the term Learning Enabled Component (LEC) used in [19]).

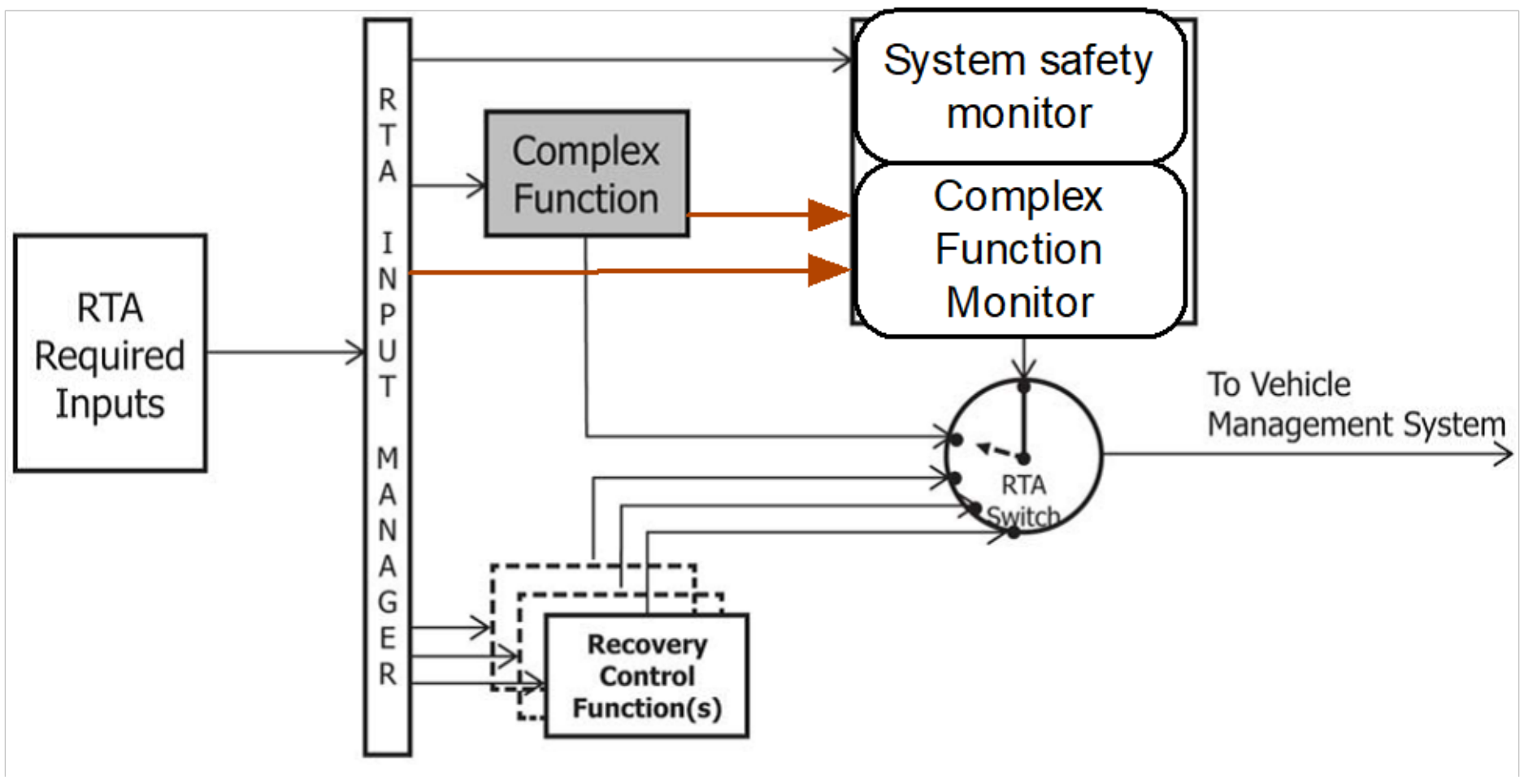

Figure 7: Safety Monitor architecture

The outline claim structure is shown in claims 3.1.1, 3.1.2, and 3.1.3 in Figure 3: there is nothing remarkable in the structure, but that the argument justifies that the system property is satisfied by the guard and the sensor. The recovery functions also must address a number of design challenges. One of the design challenges for an AI/ML monitor is illustrated in Figure 8:

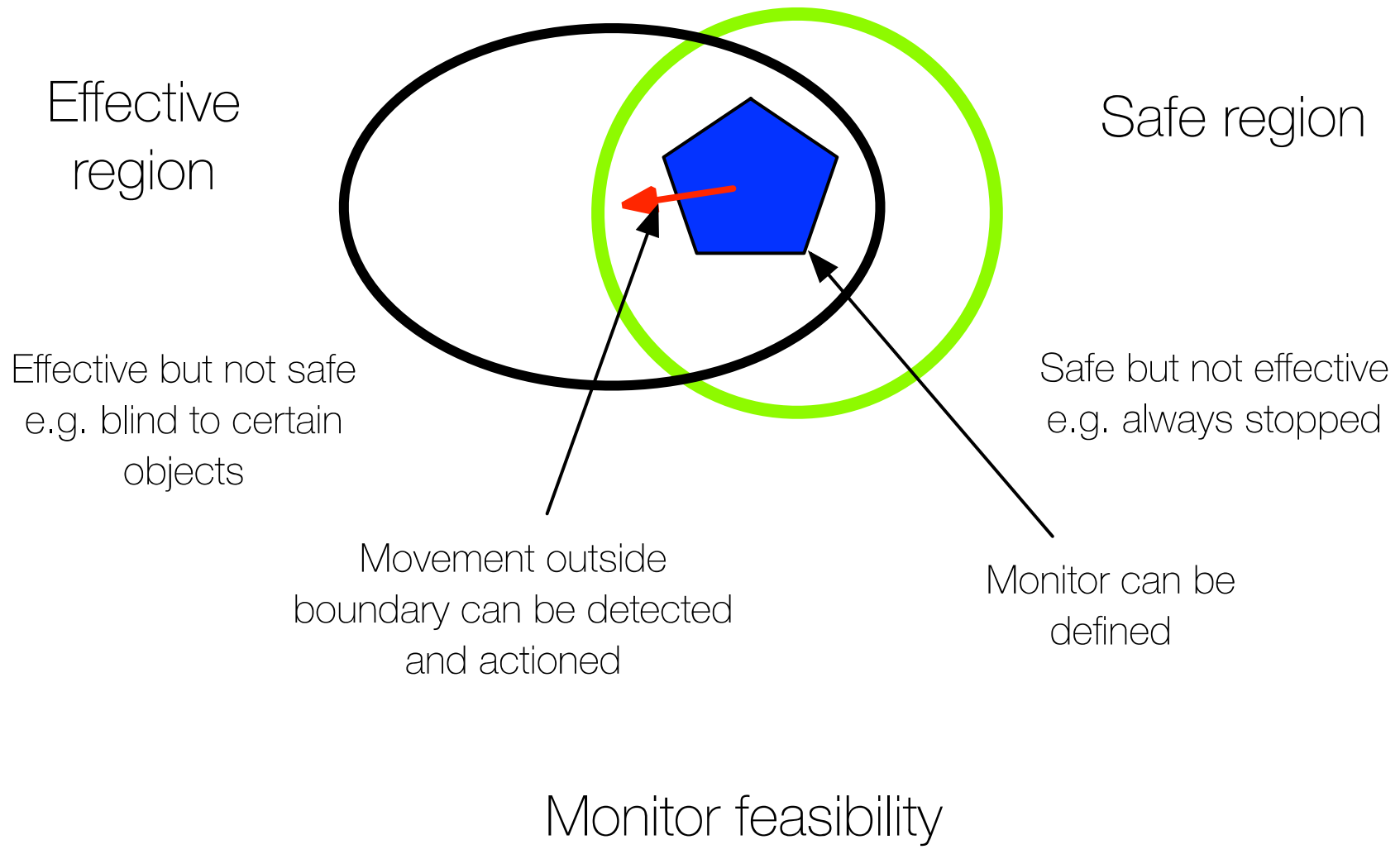

Figure 8: Monitor feasibility

In the design of this architectural strategy, a number of questions must be addressed:

- Is a monitor feasible – can a region be identified that is both safe and effective and can be described in terms that can be assured more readily and to greater extent than the AI/ML itself?
- Can transgression of the monitor region be detected and actioned by a recovery control function?

A number of approaches can then be taken, which are often characterised as

- *Environment Monitors:* Monitor an ML system's input space where there are known performance issues – e.g., bad weather, flying upside down, etc.
- *Health Monitors:* Monitor the ML system's internal state and identify states that might be "stressed" or indicative of problem, (e.g., monitoring activation patterns, resource utilisation, simple tasks for which diverse measurement).
- *Behaviour Monitors:* Monitor the ML system's outputs and inputs to see if they violate bounds on specified behaviours or invariants.

Finally, the recovery strategies are very application specific. They may involve

- Use of other sensors in the case that the ML system has degraded performance, but will still allow for safe behaviour (e.g., moving to a minimum risk position or reduced performance while recovery is planned).
- If all sensors fail have an "Eyes closed" safety strategy. For example, when solving for reachable sets in which the autonomous vehicle and other vehicles maintain safe separation and safety despite unobservability, a simple approach would be to slow down and stop.

Safety monitors are one example of defence in depth. Defence in depth is fundamental to achieving safety and resilience through detection, tolerance, recovery and disaster management and through having appropriate system architectures. Defence in depth and diversity are discussed in TTN [34].

The sensor case is based on a generic approach to component cases (see Section 2.2.1). The generic component case also highlights the need to consider how the sensor case can enable and impact other parts of the case. Some of this will be dealt with at a component level and some at a system level (see Section 2.2). In terms of the future behaviour of sensors, the case would need to address the topics identified in Section 2.3 of dealing with the change infrastructure and the change itself. These would include changes to learning tools, training data, tool chain and updates as knowledge of vulnerabilities increased. There would also be specific ML related issues around revalidation.

Assurance of the AI/ML based sensor presents challenges because of the very nature of the technology and the complex tool chains used to develop it. As with conventional software components evidence can come from dynamic analysis and static analyses and formal verification.

## 2.3 Trustworthiness in the future

As previously noted, our top-level decision concerning the deployment of the system needs confidence that the system will continue to meet its key requirements. We thus expand the claim into three aspects of the OSD process, as shown in Figure 3 claims 4.1, 4.2, and 4.3. The first two address whether the agreed changes are adequate and that they have been implemented. The third claim concerns the accountability infrastructure and whether the need for change is detected, and that the sentencing and accountability systems addressing changes are adequate. This reflects the OSD lifecycle model presented in Figure 2.

The impact of security for the different parts of the OSD process has been analysed and this shows [16] that there will be two major changes from a security perspective: the adaptation process needs to consider non-benign events and attacks on the change infrastructure itself. The change infrastructure will also need to be adapted in the face of threats and other changes so the diagram and OSD approach will be deepened, as it is applied recursively.

We note that the leaf claims in Figure 3 will require expansion and further research to address the system and its components, which will be particularly challenging for the AI and ML-based subcomponents.

# 3 Assurance framework and development of methodology

Recall that in Section 0, we note that there are significant new challenges that are associated with reasoning and evidence rather than structure. For example, extrapolating experience from simulation or trials to real use by demonstrating the reliability of ML based sensors and that training data has not been compromised, in addition to justifying the argumentation structure. While the latter can be addressed by the use of CAE Blocks, there is a need to evolve the assurance methodologies to address the other issues.

The practice of structured argumentation in assurance cases[1] has not developed drastically since ASCAD and GSN were first published in the 1990s, and in the drafting of ISO/IEC 15026, as these methods often focus on the structure of the argumentation. However, in the past 5-10 years, considerable research has been carried out on the development of assurance cases which consolidate insights from deductive reasoning and informal logic, to describe and support the reasoning that engineers make. We have dubbed this evolved approach as "Assurance2", and details are provided in an accompanying working paper [28].

[1] We use the term *assurance* case to cover cases that address different attributes: safety, security, dependability, effectiveness as well as confidence cases.

The "Assurance2" framework aims to support reasoning and communication about the behaviour and trustworthiness of engineered systems. It maintains the indication of the structure of the argumentation (as in CAE) but is strengthened with an explicit focus on the evidence and reasoning in cases. From a research point of view, we relate the conceptual framework to historical argumentation approaches from Wigmore and Toulmin [6][29], in particular, those concerning applied natural language deductivism. Our approach includes the following:

- *Reasoning explicitly and recognising inductive/deductive split*. Making explicit inference rules and the separation of inductive and deductive reasoning. The use of empirically based CAE Blocks - a restrictive set of common argument fragments - provide a mechanism for separating inductive and deductive aspects of the reasoning and for justifying the inference steps.
- *A closer examination of evidence integration*, addressing both the relevance and provenance of evidence. The concept of evidential threshold is thus introduced, allowing one to state that a claim can be reasoned about deductively.
- *Explicit use of doubts and defeaters* - both undercutting and rebuttal - to ensure that confidence is an integral part of the justification. Templates and associated narratives can be used to capture challenge in a systematic manner probing the understanding of the role of technical and human systems as well as the importance of identifying any unintended behaviours.
- *Use of Confirmation Theory* to evaluate the strength of evidence and arguments. Confirmation theory is defined as the study of the logic by which scientific hypotheses may be confirmed or disconfirmed (or supported or refuted) by evidence. When analysing a safety case, it is important to consider how compelling the evidence is with regard to the particular claim; does the evidence have the ability to confirm or disprove the claim? Does a set of sub-claims support the higher-level claim? We propose in [28] the experimental use of the Kemeny and Oppenheim [18] measure.
- *Use of bias and counter cases.* An explicit approach to reduce bias is by the use of counter-cases and confirmation theory. An interesting feature of the confirmation measure [18] we propose is that it is symmetrical in the use of claims and counter-claims: it requires the user to consider both situations where evidence supports a claim and the alternative, where evidence can act to disprove the claim.

There are additional innovations that could be deployed to support the assurance case. For example, the use of "sentencing statements" which support an individual's reasoning of the overall judgement used to justify a system's trustworthiness [28]. There is also exploratory research on concepts such as the chain of confidence (incorporated in the IAEA guide [15]) for exploring assumption doubt, as well as research into applying Bayesian frameworks for integrating judgements.

The deployment of these Assurance2 concepts is at various stages of maturity. We have currently trained 50 engineers in the concepts and we have a waiting list of 200 to be trained in 2020. We are building on Adelard's assurance case tool ASCE V5 to provide support.

# 4 Conclusions and recommendations

Most system level standards and frameworks require some form of safety case or safety justification – see for example "Safety first", UL4600, ISO26262, UK PAS 12281, UK PAS 1881. We have illustrated how such a case could be constructed, and provided a commentary on its development along with a number of technical supporting TIGARS Topic Notes on key topics such as: system lifecycles (open systems dependability perspective), resilience and requirements, security, defence in depth and diversity, simulation, and V&V (including formal verification and static analysis) of machine learning algorithms and platforms. We have shown the breadth of the issues as well as showing how a thread of reasoning could be developed from claims to evidence.

We provide overall conclusions, lessons learnt, and recommendations in [37], and hope our work will support innovation and assist others in developing assurance cases for real systems, whilst reducing the risk of developing systems that cannot be adequately assured.

# 5 Bibliography

[1] P G Bishop, R E Bloomfield, S Guerra, The future of goal-based assurance cases. In Proceedings of Workshop on Assurance Cases. Supplemental Volume of the 2004 International Conference on Dependable Systems and Networks, pp. 390-395, Florence, Italy, June 2004.

[2] T P Kelly, R A Weaver, "The Goal Structuring Notation - A Safety Argument Notation", Proceedings of the Dependable Systems and Networks 2004 Workshop on Assurance Cases, July 2004.

[3] R E Bloomfield, P G Bishop, C C M Jones, P K D Froome, ASCAD—Adelard Safety Case Development Manual, Adelard 1998, ISBN 0-9533771-0-5.

[4] P G Bishop, R E Bloomfield, A Methodology for Safety Case Development. In: F Redmill, T Anderson, (eds.) Industrial Perspectives of Safety-critical Systems: Proceedings of the Sixth Safety-Critical Systems Symposium, Birmingham 1998, pp. 194–203. Springer, London, 1998.

[5] GSN Community Standard, V2 Draft 1, May 2017.

[6] S E Toulmin, "The Uses of Argument" Cambridge University Press, 1958.

[7] L Emmet, G Cleland, Graphical Notations, Narratives and Persuasion: a Pliant Systems Approach to Hypertext Tool Design, in Proceedings of ACM Hypertext 2002 (HT'02), College Park, Maryland, USA, June 11-15, 2002.

[8] J Rushby, "Mechanized support for assurance case argumentation," in Proc. 1st International Workshop on Argument for Agreement and Assurance (AAA 2013), Springer LNCS, 2013.

[9] J Rushby, The Interpretation and Evaluation of Assurance Cases, Technical Report SRI-CSL-15-01, July 2015.

[10] R Bloomfield, K Netkachova, Building Blocks for Assurance Cases. 2nd International Workshop on Assurance Cases for Software-intensive Systems (ASSURE), International Symposium on Software Reliability Engineering, Naples, Italy, 2014.

[11] MISRA Guidelines for Automotive Safety Case Arguments V5, MISRA for public review, Nov 2016.

[12] Delong, Compositional Certification, Lecture Notes. Real-Time Embedded Systems Forum, The Open Group (TOG) conference, Toronto, Canada (2009) and the Layered Assurance Workshop (LAW).

[13] R E Bloomfield, K Netkachova, R Stroud, Security-Informed Safety: If it's not secure, it's not safe. Paper presented at the 5th International Workshop on Software Engineering for Resilient Systems (SERENE 2013), 3rd - 4th October 2013, Kiev, Ukraine.

[14] ISO/IEC/IEEE IS 15026-1:2018 Systems and software engineering – Systems and software assurance – Part 1: Concepts and vocabulary

[15] Dependability Assessment of Software for Safety Instrumentation and Control Systems at Nuclear Power Plants" (NP-T-3.27), https://www-pub.iaea.org/books/IAEABooks/12232/Dependability-Assessment-of-Software-for-Safety-Instrumentation-and-Control-Systems-at-Nuclear-Power-Plants last accessed March 2019.

[16] Security-Informed Safety: If it's not secure, it's not safe, Bloomfield (2013), R. E., Netkachova, K. & Stroud, R. Software Eng. for Resilient Systems, A. Gorbenko, A. Romanovsky, and V. Kharchenko, eds., LNCS 8166, Springer, 2013, pp. 17–32.

[17] Assurance of open systems dependability: developing a framework for automotive security and safety, Bloomfield, R. E., Butler, E. & Netkachova, K. Paper presented at the Sixth Workshop on Open Systems Dependability, 21 Oct 2017, Tokyo, Japan.

[18] J G Kemeny, P Oppenheim, "Degree of Factual Support," Philos. Sci., vol. 19, no. 4, pp. 307-324, 1952.

[19] F3269-17 Standard Practice for Methods to Safely Bound Flight Behavior of Unmanned Aircraft Systems Containing Complex Functions, ASTM International

[20] Matthew Clark, Xenofon Koutsoukos,Joseph Porter, Ratnesh Kumar, George Pappas, Oleg Sokolsky , Insup Lee, Lee Pike, A Study on Run Time Assurance for Complex Cyber Physical, AFRL/RQQA, 2013

[21] Shai Shalev-Shwartz, Shaked Shammah, and Amnon Shashua. On a formal model of safe and scalable self-driving cars. arXiv preprint arXiv:1708.06374, 2017.

[22] John Rushby. Runtime certification. In Martin Leucker, editor, Eighth Work- shop on Runtime Verification: RV08, volume 5289 of Lecture Notes in Computer Science, pages 21–35, Budapest, Hungary, April 2008. Springer-Verlag

[23] Safety first for automated driving. Accessed December 2019. https://www.daimler.com/documents/innovation/other/safety-first-for-automated-driving.pdf.

[24] High-Level Expert Group on Artificial Intelligence, Ethics guidelines for trustworthy AI, available online (https://ec.europa.eu/digital- single-market/en/high-level-expert-group-artificial-intelligence).

[25] Shaping the ethical dimensions of smart information systems – a European perspective (SHERPA), Deliverable No. 3.2. BSI ART/1_19_0257

[26] Uber ATG. Safety Case. Accessed December 2019. https://uberatg.com/safetycase.

[27] Pegasus Projekt. Accessed December 2019. https://www.pegasusprojekt.de/en/home.

[28] R Bloomfield et al, A new framework for CAE based assurance cases and engineering justifications, Adelard W/3013/138008/19, Dec 2019.

[29] Wigmore John Henry, The Science of Judicial Proof, Vol. 25, No. 1 (Nov., 1938), pp. 120-127, Published by: Virginia Law Review. DOI: 10.2307/1068138.

[30] TIGARS Topic Note, Resilience and Safety Requirements, D5.6.2 (W3035). December 2019.

[31] TIGARS Topic Note, Open Systems Perspective, D5.6.3 (W3036). December 2019.

[32] TIGARS Topic Note, Formal Verification and Static Analysis of ML Systems, D5.6.4 (W3014). December 2019.

[33] TIGARS Topic Note, Simulation and Dynamic Testing, D5.6.5 (W3015). December 2019.

[34] TIGARS Topic Note, Defence in Depth and Diversity, D5.6.6 (W3021). December 2019.

[35] TIGARS Topic Note, Security-Informed Safety Analysis, D5.6.7 (W3022). December 2019.

[36] TIGARS Topic Note, Standards and Guidance Relevant to Assurance of Autonomy, D5.6.8 (W3025). December 2019.

[37] TIGARS Summary and Recommendations, W3033. December 2019.

# TIGARS

Towards Identifying and closing Gaps in Assurance of autonomous Road vehicleS

**Project Ref: 01/18/05**

**Adelard Ref: W/3025/138008/27**

# TIGARS TOPIC NOTE 2: RESILIENCE AND SAFETY REQUIREMENTS

## Summary

This TIGARS Topic Note discusses resilience analysis and safety requirements for autonomous vehicles. We provide a background to resilience analysis and use the Open Systems Dependability lifecycle to help in defining safety requirements for an example autonomous service.

24 Waterside
44–48 Wharf Road
London
N1 7UX

T +44 20 7832 5850
F +44 20 7832 5870
E office@adelard.com
W www.adelard.com

Authors
Yutaka Matsubara
Robin Bloomfield

## Use of Document

The document is made available as a resource for the community, providing all use is adequately acknowledge and cited.

This document provides a snapshot of work in progress. We welcome feedback and interest in this work. Please contact the authors or admin.tigars@adelard.com

Acknowledgement

This project is partially supported by the Assuring Autonomy International Programme, a partnership between Lloyd's Register Foundation and the University of York. Adelard acknowledges the additional support of the UK Department for Transport.

## Contents

## Tables

## Figures

# 1 Introduction to resilience

## 1.1 Definition of resilience

A resilient system is an adaptive system, one that responds to change, and that can survive and prosper when challenged: a system that can deal with attack and can deal with surprises. It is common to think of resilience in terms of the stimulus and recovery model shown in Figure 1.

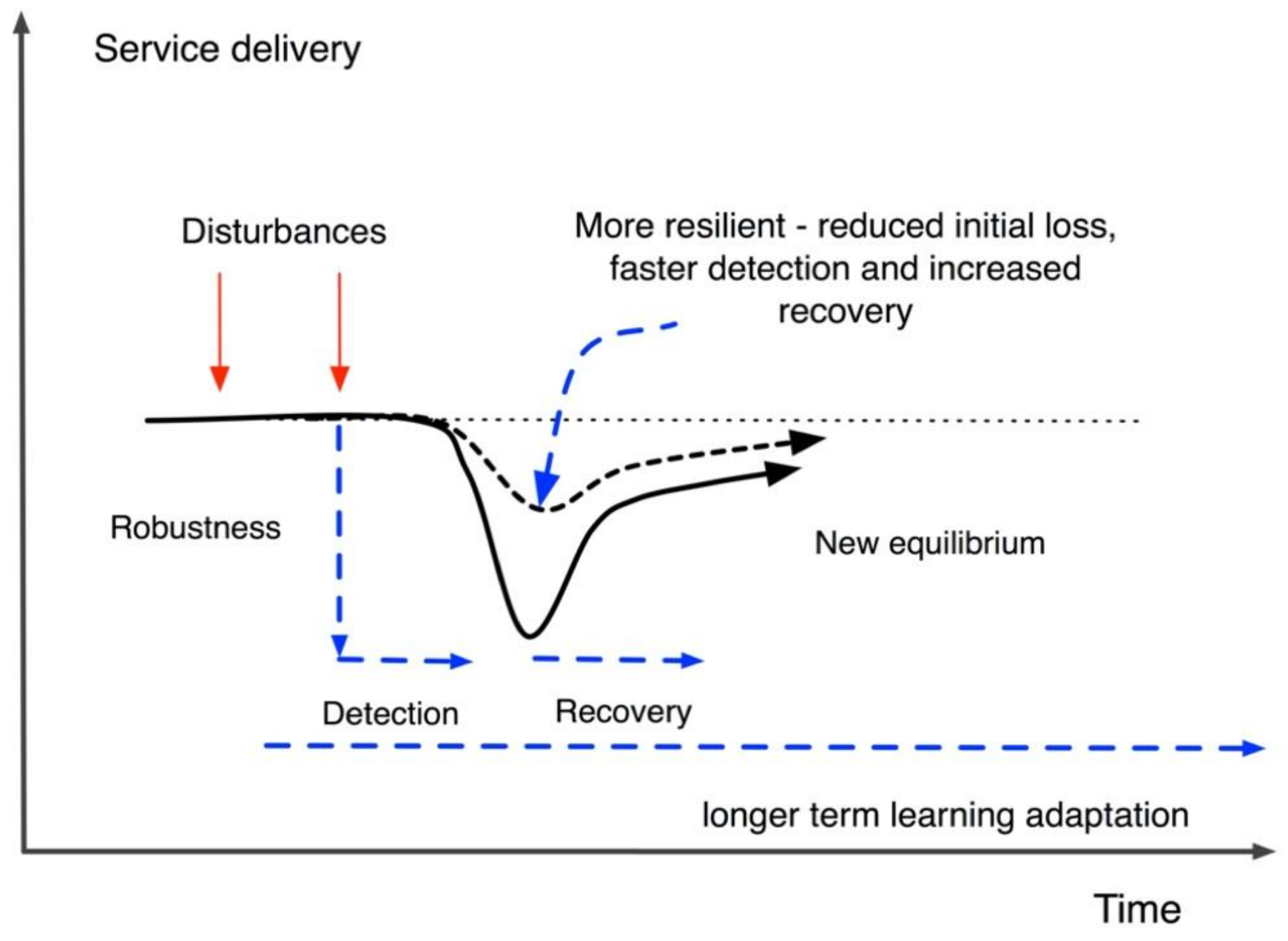

**Figure 1: Resilience**

In the last decade there has been much work on developing the concept of resilience, drawing on psychology and social science as well as the traditional engineering disciplines [1]. The emphasis is on the ability of a system to adapt and respond to changes in the environment.

We find it useful [1][2] to distinguish two types of resilience:

- Type 1: resilience to design basis threats and events. This could be expressed in the usual terms of fault-tolerance, availability, robustness, etc.
- Type 2: resilience to beyond design basis threats, events and use. This might be split into those known threats that are considered incredible or ignored for some reason and other threats that are unknowns.

Often, we are able to engineer systems successfully to cope with Type 1 resilience, but Type 2 resilience is a more formidable challenge.

Furthermore, resilience is more than just the ability to cope with external stimuli. Complex systems can be challenged not just by exogenous events but also by those internal to the system: some of these might be the traditional failure of components (either technical or human). However, increasingly, it is claimed, significant failures are due to an accumulation of normal variability [1]. Furthermore, these significant failures in some instances become correlated, and because of the very non-linear response of the system, they result in unexpected and/or unwanted behaviours. In this last scenario, reductionism is a much less successful strategy and a more holistic approach is necessary or even essential.

This document considers the use of resilience models and open system dependability (OSD), as to keep and improve service-level dependability.

OSD targets open systems, which is a system constantly changing its structure and boundaries between internal objects and external ones. In this situation, it is difficult for one stakeholder to understand an entire open system. IEC 62853 specifies the four perspectives of response, change response, consensus building, and accountability [8]. To continuously plan and monitor, appropriate responses are required throughout the system life cycle to maintain and continue dependability as much as possible.

In this research, we have focused on the recoverability and adaptability of a service, and defined resilience as *"the ability of the service to adapt to external and internal changes so that functionalities in the service can be continued as much as possible or be recovered as fast as possible"*.

Based on these perspectives, there are a number of key characteristics to consider when addressing resilience: our epistemic uncertainty, the endogenous/exogenous nature of the events and the extent to which they are emergent system properties or addressable by more reductionist approaches. There are many research challenges to be addressed, and in the short term there are potential gains from understanding the different insights from the "normal accidents" school [3], the High Reliability Organisations (HRO) protagonists [4], and the resilience and socio-technical dependability communities [5][6][7].

## 1.2 Automotive standards and trends

Modern vehicles have ADAS (Advanced Driving Assistance Systems) and/or automatic driving functions to improve their safety. There is intense competition to develop autonomous vehicles, which includes driver-less vehicles. As systems become more automated and autonomous, safety-related functions are increasing in scale and complexity. Most of these are realized by electrical and electronic systems with software.

Functional safety standards ISO 26262 and ISO/PAS 21448 have been published for electrical, electronic and programmable systems related to safety. ISO 26262 mainly aims to analyze and take safety measures against the impact of malfunctions caused by failures in the system. The purpose of ISO/PAS 21448 is also to analyze and take safety measures against the performance limitations of sensors and AI (Artificial Intelligent) components, and the misuse of drivers regarding HMI (Human Machine Interface). ISO/PAS 21448 is currently published as a PAS (Publicly Available Specification) for automated driving with levels up to 2. The discussions on ISO 21448 (International Standard) for autonomous driving with level 3 and above have already begun. Regarding automotive security, discussions are also proceeding towards the publication of draft of ISO/SAE 21434 in 2020.

## 1.3 From vehicle-level safety to service-level safety

Current international standards have discussed the safety of the vehicle itself i.e. system safety. However, this does not address Mobility as a Services (MaaS), where autonomous driving systems and/or AI components may be used. MaaS is a concept of a service that provides optimal transportation by the means of appropriately combining kinds of transport, which includes not only cars but also bicycles, buses, trains, airplanes, etc., with respect to users' requests.

For MaaS, social infrastructures and people such as users, mobility, and urban transport, traffic sign, etc. are widely connected with each other. When considering the safety of users, it is not enough to achieve vehicle-level safety. For example, if a critical failure happens in a vehicle, vehicle-level safety can usually be achieved by slowing the vehicle down and stopping it on the shoulder of the road even if the vehicle is running in a dangerous zone. For service-level safety, we need to manage and maintain the service, even when a vehicle may have performed an emergency stop due to failures, by arranging alternative transport to achieve their purpose and recovering the failed vehicle to reduce the time at risk. To support that, it is also necessary to provide status information about the stopped vehicles to all stakeholders including related users, service managers, and vehicle maintainers in order to understand the situation, to make decisions according to their purpose or desire, and to avoid secondary accidents.

As mentioned earlier, in the international safety standards for E/E/P systems for automotive, the hazardous factors considered are limited to vehicle-level malfunction. To achieve the service level safety or more general safety defined in ISO/IEC Guide 51 as 'freedom from risk which is not tolerable', it is necessary to remove or mitigate unacceptable risk from the user's point of view, which means that not only vehicle-safety but also dependability is important to support adequate quality of services, and includes convenience, reliability, durability, maintainability and security. Unfortunately in the automotive field, there are no standards that focus on dependability or service-level safety. In ISO 21448, a safety mechanism for continuously monitoring and updating vehicle components (especially sensors and AI) after product release is under discussion. ISO AWI 22737 (Low-Speed Automated Driving Systems) is being developed to standardize an evaluation method for the safety and performance of services for autonomous driving vehicles moving at low speeds. This is just at Preliminary work item (PWI) stage, and detailed requirements will be discussed in the future.

### 1.4 Resilience Analysis and Evaluation

The goals of resilience and Open System Dependability (OSD) are similar; however, the methodologies and evaluation methods for achieving their goals are not well established. For OSD, guidelines based on ISO/IEC/IEEE 15288 (Systems and software engineering-System life cycle processes) are specified in IEC 62853 [8]. However, since it is a generic standard, it is necessary to interpret or refine the requirements according to the target service or system to which it is applied. In addition, detailed requirements presented as outcome, activities and tasks in IEC 62853 can be defined in accordance with ISO/IEC/IEEE 15288, but their number is enormous. It is relatively easy to apply to a system if it has been well developed based on ISO/IEC/IEEE 15288; but it is not easy to apply it to the automotive field at this time.

For the qualitative and quantitative evaluation for resilience, there have been academic studies, but no international standards exist. We have applied System-Theoretic Process Analysis (STPA) to a FRAM model [9] in our approach. Through the trial, we have confirmed that STAMP/STPA and FRAM/STPA are suitable to analyse hazards caused by resonances or relationships among components including AI functions. Additional requirements for a service level analysis method are also implied in order to make service level requirements clearer before system/component level analysis.

In this research, for the purpose of maintaining and improving the dependability of MaaS with AI components, the functional requirements and performance requirements are considered, specified and verified repeatedly at the service development phase. To develop an assurance case of resilience, we propose a new qualitative evaluation method based on IEC 62853. We used the Goal Structure Notation (GSN) templates in Appendix of IEC 62853 and adapted them to create an assurance case for resilience of the target service (see Appendix A). Using a GSN argument in the development phase has the advantage that service providers can understand which requirements have been already met and those which have not using the service specifications. If critical requirements for resilience are not met, new specifications regarding functionality or countermeasures should be added or modified at the early development phase, *resilience by design*.

## 2 Resilience analysis based on IEC 62853

### 2.1 Resilience Analysis Work Flow

The proposed resilience analysis work flow is shown in Figure 2. The details of each step are described as follows.

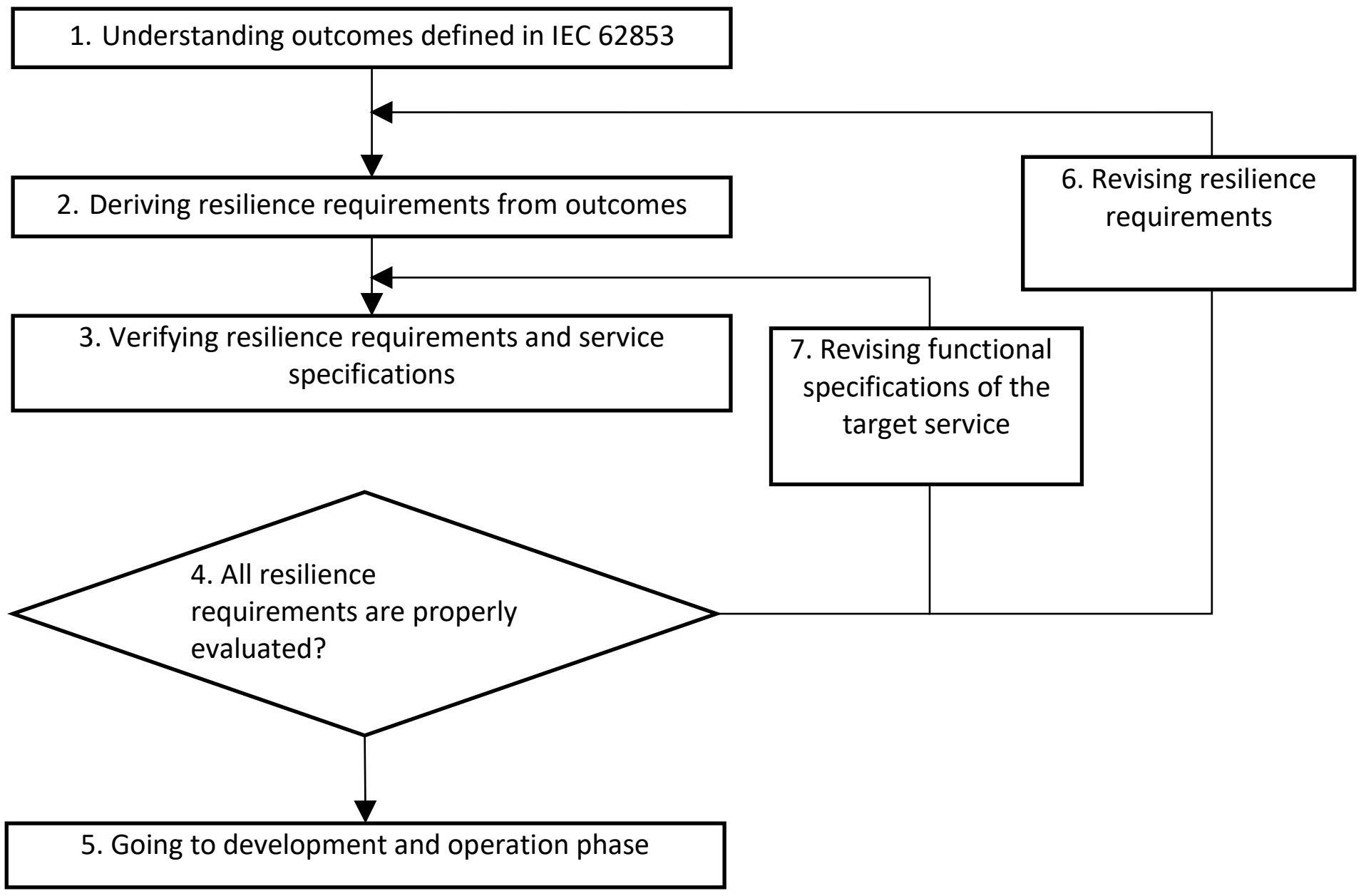

Figure 2: The resilience analysis flow based on IEC 62853

1. Understanding outcomes defined in IEC 62853

The outcomes to be achieved are defined for each of the four OSD process views of IEC 62853. In order to realize these outcomes, detailed requirements are defined as process activities and tasks specified in ISO/IEC/IEEE 15288. Prior to analysis, we understand the intended meaning of outcomes/results. In this study, we focused on the failure response process view that is especially related to resilience.

2. Deriving resilience requirements from outcomes in IEC 62853

The outcomes in IEC 62853 are abstract so that they can be applied to services in kinds of domains. We, service providers, need to clarify and specify resilience requirements such as uptime and downtime, and damage to be avoided in our service. The outcomes and resilience requirements are presented as *goal* nodes in the GSN argument.

3. Verifying resilience requirements and service specifications

The verification results, which include whether each resilience requirement can be met or not by the current service specification, are presented as *evidence* nodes in GSN argument. If an outcome or a resilience requirement presented as *goal* node can be met, related documents are specified as *evidence* node. If not, the reason why it cannot be satisfied at this phase or when it will be satisfied at a later phase is described in the supporting *evidence* node.

4. Evaluating resilience requirements

After step 3, we evaluate whether each outcome in IEC 62853 is satisfied. Where outcomes are not satisfied this is explained in the related *evidence* node. If all outcomes are met or correspond to appropriate justifications, we can go to step 5. If there are unsatisfied outcomes without valid justifications, we go to step 6 or 7.

5. Going to development and operation phase

After resilience requirements and service specifications are verified and validated respectively through steps 1,2,3,4, 6 and 7, then the target service can proceed to development and operation phases. In the operation phase, resilience requirements, which have been already satisfied at the design phase and development phase, are continuously monitored to see whether they are still satisfied.

6. Revising resilience requirements

If we realise that any requirements derived from outcomes at the step 2 are incomplete or inappropriate, they should be revised or new resilience requirements will be added.

7. Revising functional specifications of the target service

To meet resilience requirements, functional specifications of the target service will be revised or added.

## 2.2 Target Service for the case study

The target service is a fictional automatic package transportation service as shown in Figure 3. Resilience analysis and design support will be implemented for this service based on the requirements of the failure response process view of IEC 62853. This automatic transport service starts to be used when a user puts a package on an autonomous driver-less transport vehicle and issues a transport request to the operator via a smartphone. When the operator accepts the transportation request and issues a transportation instruction to the autonomous driving vehicle, then transportation starts. When the transport is completed successfully, the recipient is contacted by an application on the smartphone, and the recipient completes by taking the package from the parked vehicle. If the vehicle stops en-route e.g., due to unexpected failures during transportation, the operator tries to restart the system of the vehicle. If it still does not recover, the operator arranges an alternative vehicle with vehicle maintainers and resumes the transportation service.

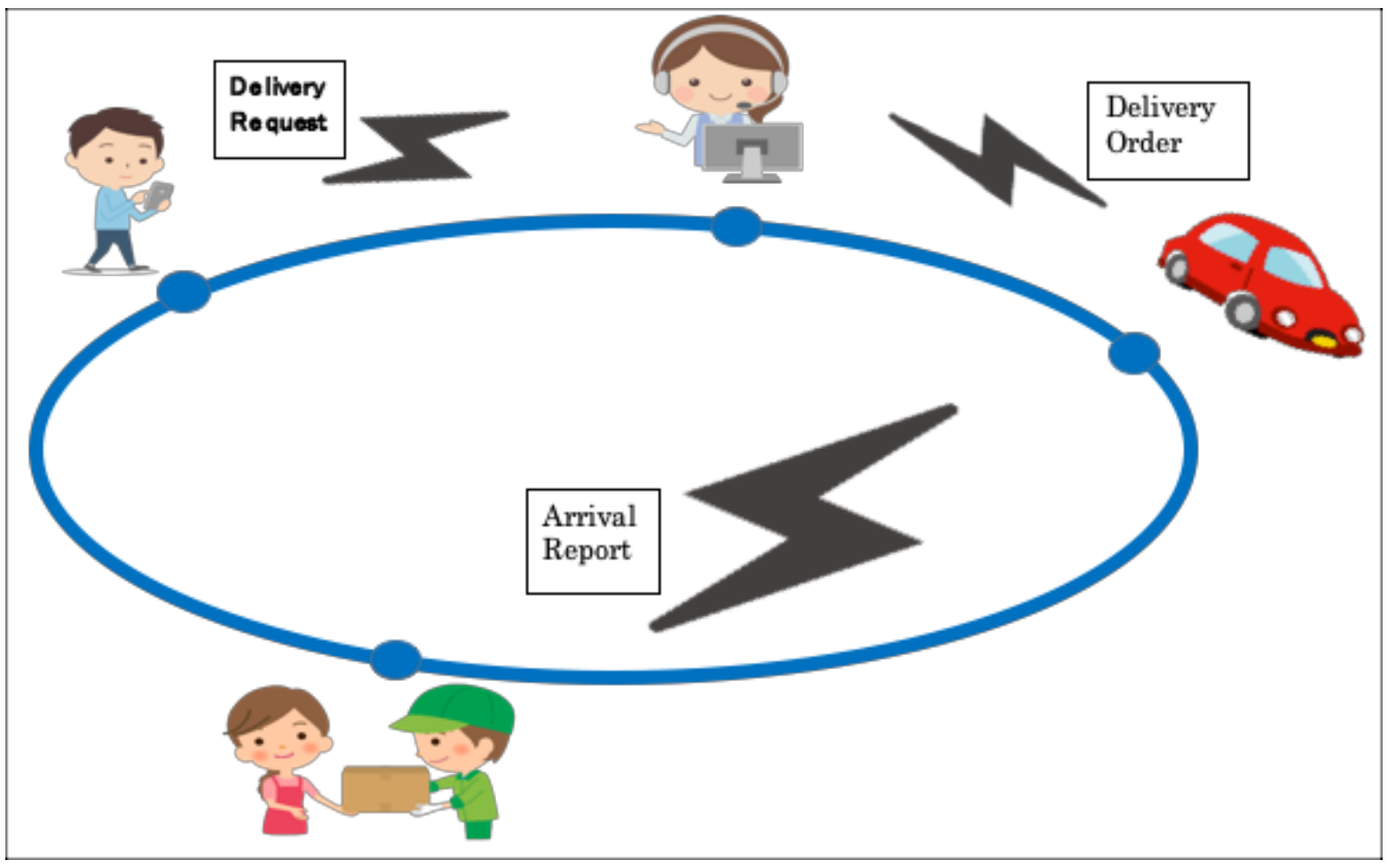

**Figure 3: Overview of the automatic package transportation service**

The following specifications were developed as the preliminary (and initial) specifications for the services. An ID, which is presented as 'SS-XXXX', is assigned to each specification (see Table 1). 'SS' stands for 'Service Specifications'. 'XXXX' means a number of service specifications. SS-0051 and SS-0052 are the detail specifications derived from SS-005.

Service providers usually focus on what they want to provide as the service. The preliminary specifications can be often ambiguous and incomplete from the perspectives of safety and resilience since developers have limited experience and expertise especially in case of MaaS. The proposed analysis improves the specifications so that resilience requirements are satisfied as possible as much from the service design phase.

| ID | Specification description |
|---|---|
| SS-001 | A sender(customer) puts(loads) a package on a vehicle, and a receiver takes(unloads) it manually. |
| SS-002 | The sender requests its delivery to the transportation system by his/her smart phone after completed loading. |

| ID | Specification description |
|---|---|
| SS-003 | The transportation system orders starting delivery to a vehicle (an autonomous car). |
| SS-004 | The vehicle transports automatically the package to the destination. |
| SS-005 | The vehicle reports the result of transportation after completed it. |
| SS-0051 | The vehicle reports the success status to the receiver if the delivery was completed successfully. |
| SS-0052 | The vehicle reports the fail status to the receiver and the operation department if the delivery was failed due to serious reasons. |
| SS-006 | The operation department takes adequate responses, which include rebooting the control system of the vehicle and arranging an alternative vehicle, according to the situation when any troubles happen during transportation. |
| SS-007 | The operation department requests the development department to make corrections if the cause is serious. |
| SS-008 | The development department fixed and/or modifies service specifications if needed. |

**Table 1: System specification**

## 2.3 Case Study

The preliminary model which includes the initial service specifications is shown in Figure 4. During the development of the model, we realized that several important functions were lacking in the initial specification, and it was updated and improved. We started the resilience analysis based on Figure 2.

To understand the service specifications from all stakeholders of the services, we have used a form of FRAM model. Each function of the service is presented as a coloured square, and has inputs (I) and outputs (O). All outputs are connected to functions as these inputs with white boxes, which describes data sent from output to input. Several functions may have controls (C), resources (R) and/or preconditions (P). Please refer to [9] for more details. The colour of each function in Figure 4 represents which stakeholders "own" the function. For example, the car manufacturer has four functions coloured yellow. Through the development of the FRAM model, all stakeholders can easily understand the service specifications.

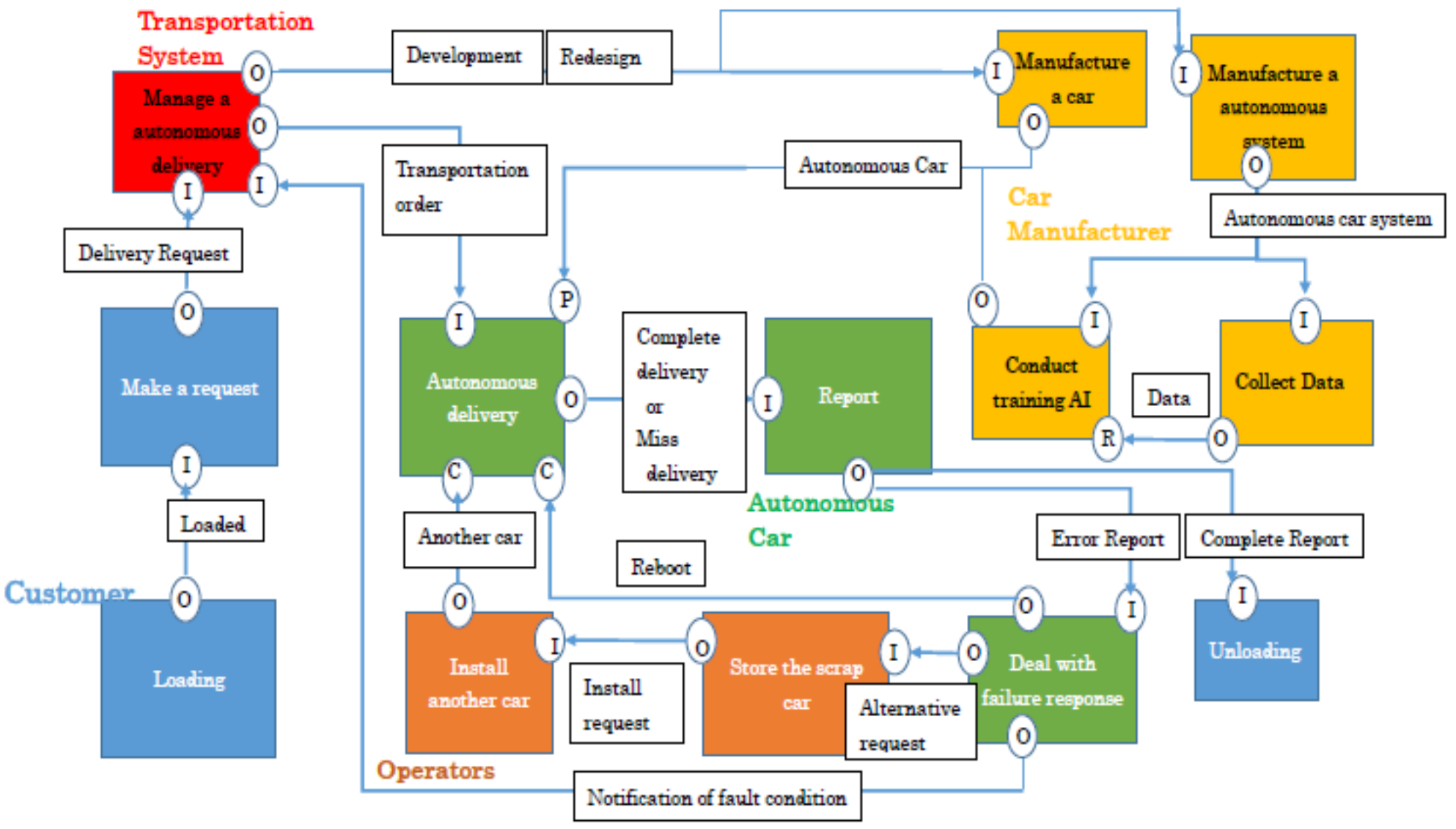

**Figure 4: The initial model by FRAM**

We started the resilience analysis with the FRAM model shown in Figure 4. We followed step 2 in Figure 2 and interpreted the outcomes defined in the failure response process view in IEC 62853 to specify resilience requirements as follows. The ID corresponds to nodes in the GSN argument structures.

| ID | Outcomes in IEC 62853 | Resilience requirements for the target service |
|---|---|---|
| G0 | The Failure Response process view is achieved. | The Failure Response process view of the target service is achieved. |
| G1 | The provision of the service is continued as much as possible, with the least possible disruption and damages, in the manner most expedient in the context. | The provision of the autonomous delivery service is continued as much as possible, with the least possible disruption and damages, in the manner most expedient in the context. |
| G2 | Immediate harms of failures are mitigated. | When the autonomous delivery service is stopped, recovery procedures such as resumption of delivery and compensation) are performed for users. |
| G3 | Longer-term harm of failures is mitigated: public confidence in the system and continual improvement are sustained. | The long-term inability of the service is reduced. Trust and continuous improvement of automatic transportation services are sustained. |
| a1) | Key functions to be protected in order to ensure service continuity are identified. | Key functions to be protected in order to ensure autonomous delivery service continuity are identified. |
| a2) | Goals for protection of the key functions necessary for continuous provision of service are identified. | Goals for protection of the key functions necessary for continuous provision of the autonomous delivery service are identified. |
| a5) | For the identified faults, errors, failures and their precursors, the goals of treatment necessary for continuous provision of service are defined and agreed. | For the identified faults, errors, failures and their precursors, the goals of treatment necessary for continuous provision of the autonomous delivery service are defined and agreed. |
| a7) | Specific responses that protect the key functions from faults, errors, failures and their precursors in class a)6)i) and default responses to those in class a)6)ii) and a)6)iii) are developed. | A specific response process is performed for a countermeasure for faults, errors, and failure expected when they are detected, and a default process is performed for a countermeasure for faults unexpected. |
| c2) | Confidence and trust in the system is sustained. | Confidence and trust in the autonomous delivery service is sustained. |
| c3) | Stakeholders and society in general are informed of the account of the failure response. | Users, developers and maintainers are informed of the account of the failure response. |

**Table 2: Resilience requirements**

Through the loop of step 3, 4, 6 and 7, we have obtained the improved FRAM model as shown in Figure 5. This model includes more than additional 20 specifications for the target service.

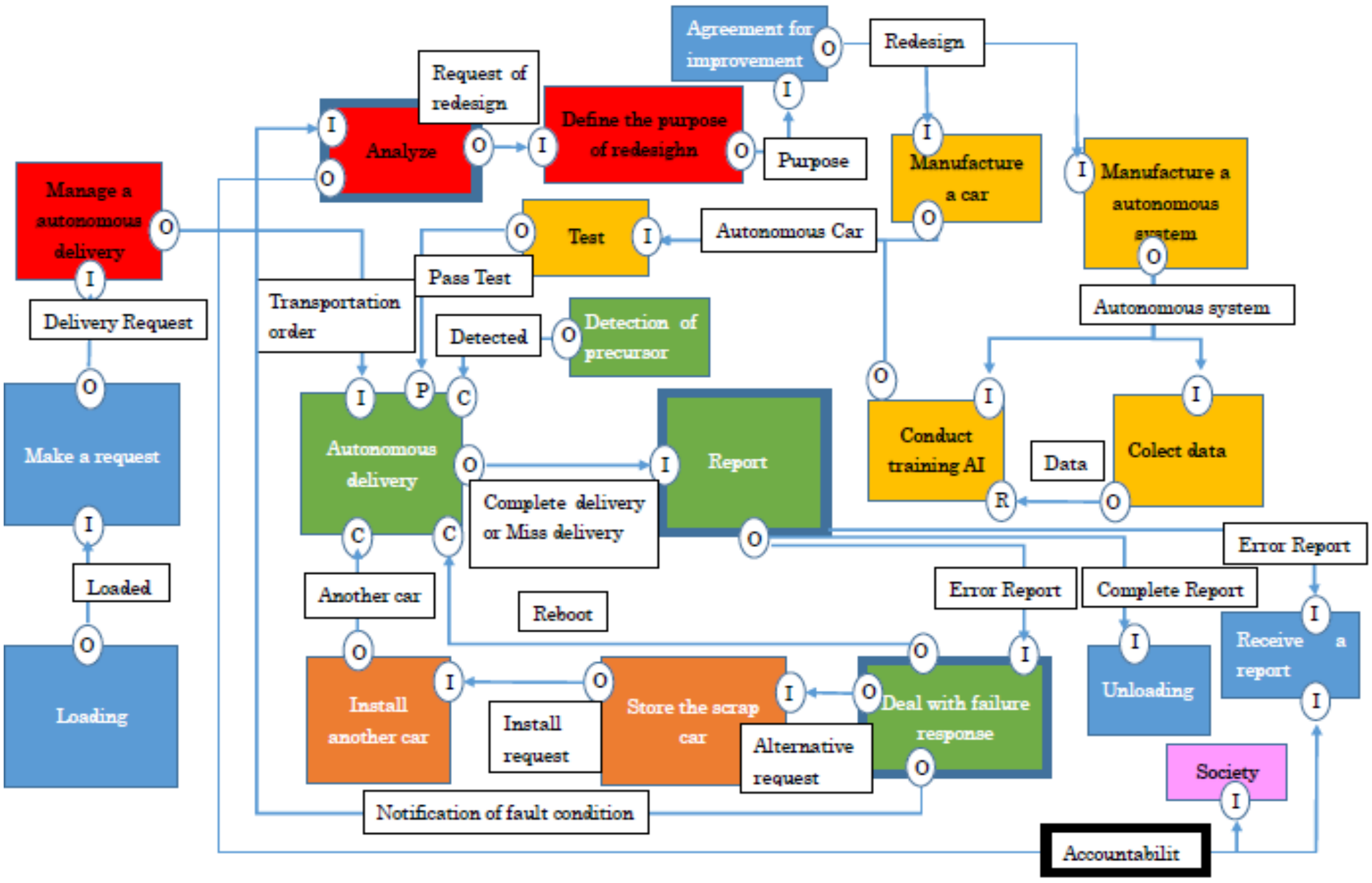

Figure 5: The improved FRAM model through the proposed analysis

The final results of this resilience analysis are presented as GSNs shown in Figure 6 to Figure 11 (see Appendix A for the more detailed Figures Figure 7 to Figure 11). Table 3 explains the colours used to indicate status of the individual goals.

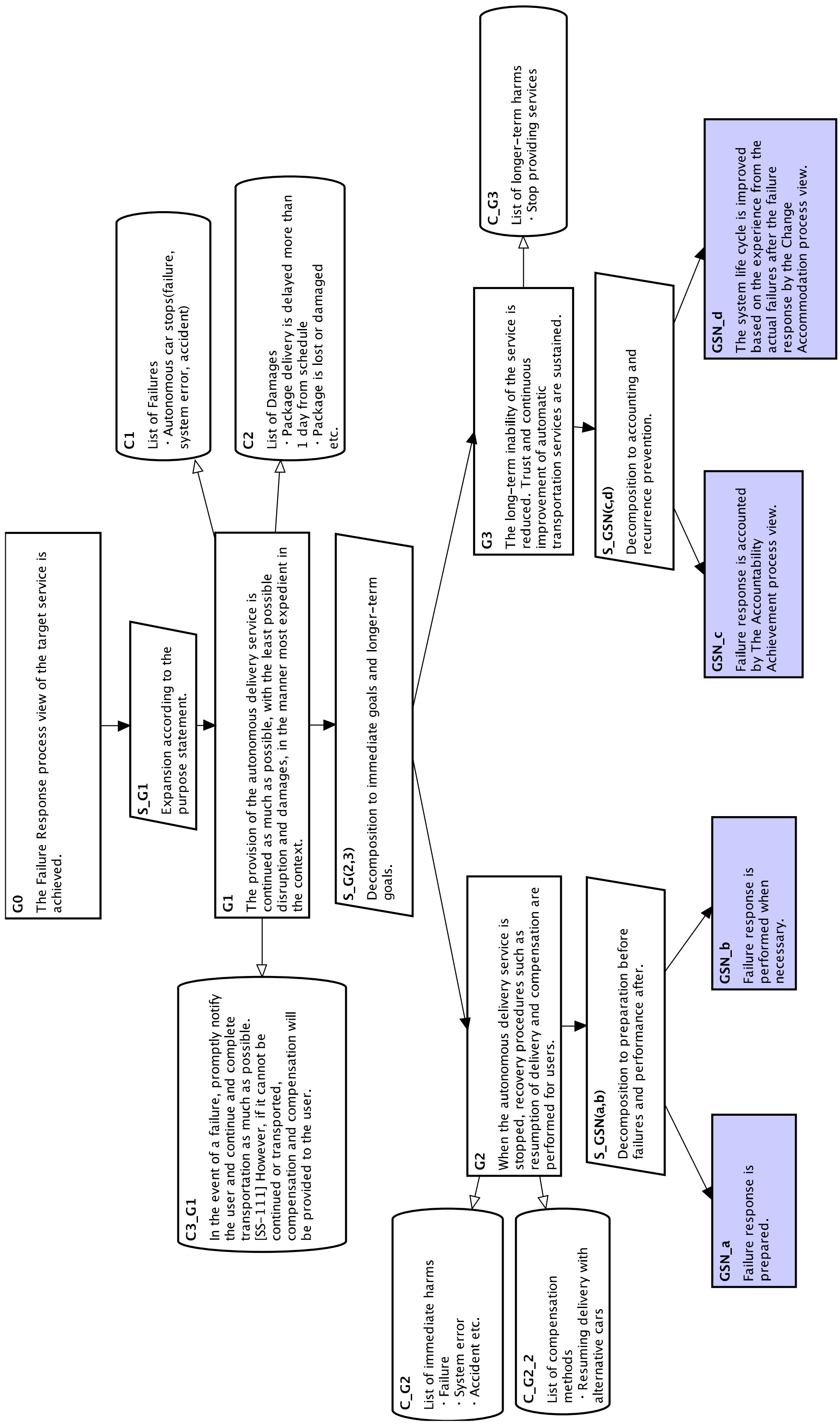

Figure 6: The top of GSN for the failure response process view

| Colour | Analysis results |
|---|---|
| White | The original outcomes in IEC 62853 or the resilience requirements derived from outcomes. |
| Purple | The goal is expanded to another detail GSN. |
| Green | The current specifications can completely meet the goal. |
| BB Orange | The current specifications can partially meet the goal. Further development and/or consideration are/is required. |
| Yellow | The goal cannot be satisfied by the current specifications only, but it may be able to be partially or fully satisfied when it can be assumed that the service or the vehicle is developed according to other safety standards (ISO 26262 or ISO/PAS 21448 etc.). |
| Red | The current specifications cannot meet the goal. Further development is required, or this goal should be evaluated at operation phase or more. |

**Table 3: GSN node colours**

## 2.4 Summary of resilience analysis

Following steps 3, 4, 6 and 7, we developed improved FRAM model and service specifications, which meant that the proposed analysis method was found to be effective as a way to improve the service's resilience. However, as shown in the figures, some requirements cannot be met by the specifications due to the following reasons.

- In Figure 7, to completely meet a1), a2) and a8) - more detailed hazard analysis is required at the service level. In this research, only the preliminary (brainstorming-based) analysis was performed.
- In Figure 7 and Figure 8, to meet a3), a4, a5) and a6) - more detailed hazard and risk analysis and considerations for resilience measures are required as well, at not only the service level but also at the vehicle level; this means that the safety development activities specified in ISO 26262 and ISO/PAS 21448 at vehicle level will be useful and effective to meet them.
- In Figure 9, b3) and b8) - cover flexibility of failure response. Therefore, these should be evaluated through the whole of life cycle, especially during the operation phase.
- c4) in Figure 10 and d2) in Figure 11 - regarding the provision of necessary information to stakeholders, these requirements should be validated by not only the failure response process view but also the accountability achievement process view and change accommodation process view, respectively. As this research covered only failure response process view, these were not considered in depth.
- c2) in Figure 10 – regarding the sustainability of confidence and trust in the service, this requirement should be also evaluated through the whole of lifecycle, especially at operation phase.

# 3 Lessons learnt and problems

## 3.1 Guidance and recommendations

This research has contributed to the following areas:

- A novel methodology focusing on IEC 62853 failure response process view to improve service level dependability and resilience was proposed.
- The overall design and analysis flow, which can be applied to not only the vehicle level design but also the service level design, was presented.

- The proposed methodology was applied to a fictitious automated baggage transport service, and its effectiveness was confirmed.
- By applying this methodology at the target service design stage, the all resilience requirements to be monitored and verified at the operation stage were clarified.
- As a part of analysis, the relationship between service level resilience and the automotive international standard such as ISO 26262 and ISO/PAS 21448 activities were considered. These are presented as yellow-coloured goals and solutions in the GSN argument.

We make the following recommendations for resilience and safety requirements analysis:

- For dependability or resilience of the service, discussion and argumentation for that should be started from service level, not from systems or components level including vehicles, infrastructure sensors, cloud systems etc. Resilience requirements should be derived at the service level, and then assigned to each system or component.
- Machine Learning is just one method to realize functional requirements at system or component level if it is used for object recognition or decision making in a vehicle control system. Therefore, where functional and/or performance requirements are assigned to the ML-system (or ML-components), impact analysis of such as failures, error, faults and security threats analysis should be discussed from the system level. The analysis results should be fed back to the service level view.
- According to the characteristics of the AI, safety architecture is required and should be designed using checking output of AI component, detecting misjudgement by the monitoring system, and redundant/diverse components; such safety architecture can be considered at both of system/component and service levels.
- The performance of ML components may change depending on the environment. The new requirements assigned to not only the component but also other systems or components in the service may also be required, such as monitoring and verification of the behaviour of the ML components in operation monitoring and re-learning and update of the ML components. Due to optimization, additional requirements should be discussed and specified at service level not at system or component level.
- Based on the above, requirements, specifications, and verification and validation at the service level and at the system level are relevant to each other in whole of lifecycle of the service. ML should be maintained continuously during lifecycle.

## 3.2 Future work and issues to address

The issues to be solved in future are as follows.

- Other process views in IEC 62853, change accommodation process view, accountability achievement process view and consensus building process view should be also combined to the resilience analysis method.
- Discussion about clear relationships between the existing international standard, ISO 26262, ISO/PAS 21448 and ISO/AWI 22737, and lifecycle of service.
- Application of the proposed resilience analysis method to real system development based on ISO 26262 and ISO/PAS 21448.

# 4 Bibliography

[1] E. Hollnagel, D. D. Woods, and N. Leveson, Resilience engineering: Concepts and precepts. Aldershot, UK: Ashgate, 2006.

[2] R. Bloomfield and I. Gashi, Evaluating the resilience and security of boundary-less, evolving socio-technical systems of systems, DSTL research report, Centre for Software Reliability, City University London, 2008. http://www.csr.city.ac.uk/people/ilir.gashi/Papers/2008/DSTL/

[3] R. E. Bloomfield, N. Chozos, K. Salako: Current Capabilities, Requirements and a Proposed Strategy for Interdependency Analysis in the UK. In Critical Information Infrastructures Security, 4th International Workshop, CRITIS 2009, Bonn, Germany, September 30 - October 2, 2009. Revised Papers Springer LNCS 6027, 2010: 188-200.

[4] C. Perrow, Normal accidents: living with high-risk technologies, New York, Basic Books, 1984.

[5] G. Rochlin, "Defining High Reliability Organisations in Practice: a Taxonomic Prologue", in New Challenges to Understanding Organisations, Macmillan, 1993.

[6] G. Baxter and I. Sommerville, "Socio-technical systems: From design methods to systems engineering", Interacting with Computers, 2010.

[7] M. Tokoro, "Open Systems Science, from understanding principles to solving problems", IOS Press, ISBN 978-1-60750-468-9, 2010.

[8] IEC, "International Standard Open systems dependability", Edition 1.0, 2018.

[9] Y. Toda, Y. Matsubara and H. Takada, "FRAM/STPA: Hazard Analysis Method for FRAM Model", FRAMilly, 2018.

# Appendix A
## Resilience analysis diagrams

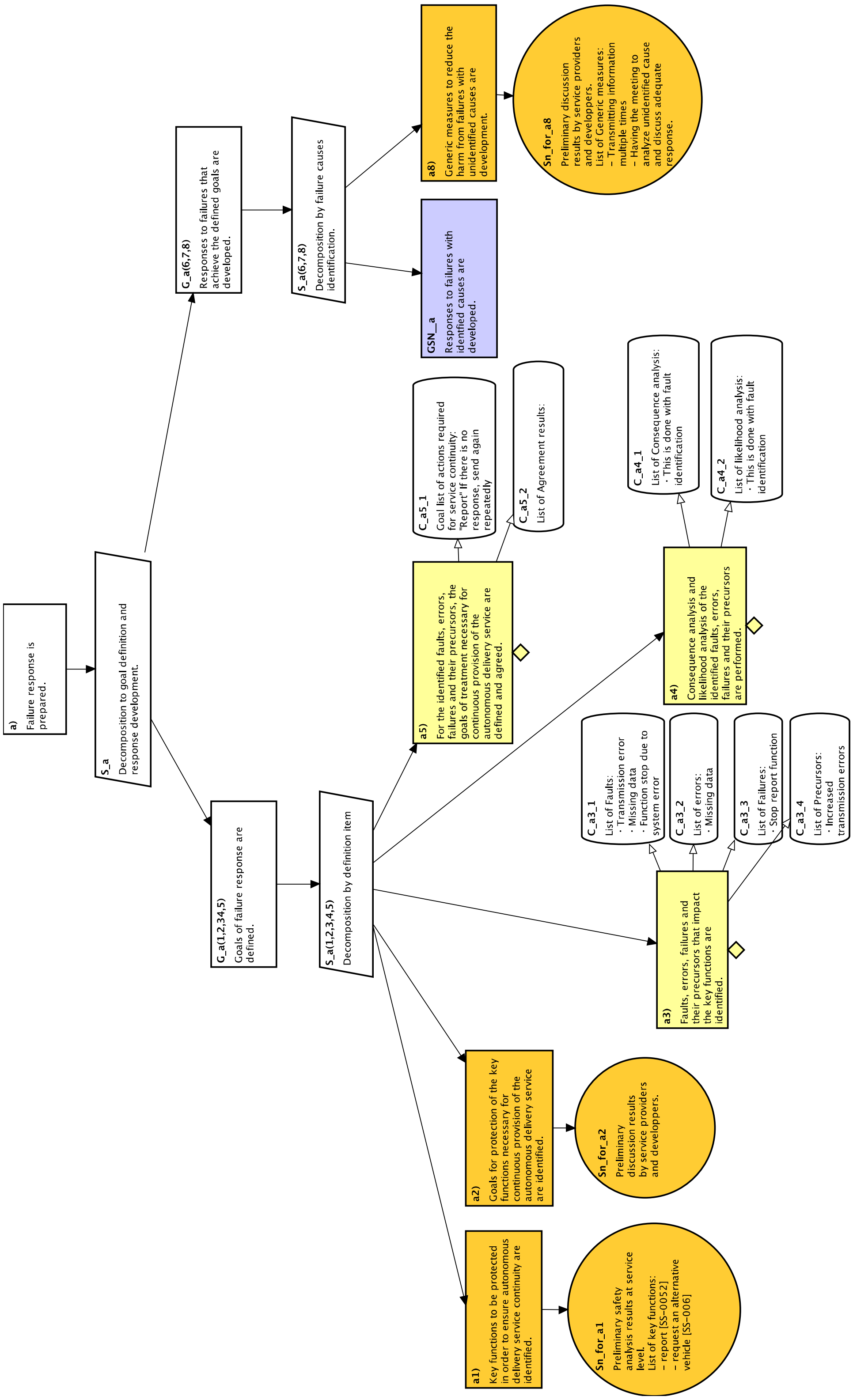

**Figure 7: The GSN for the outcome of 'a) failure response is prepared.'**

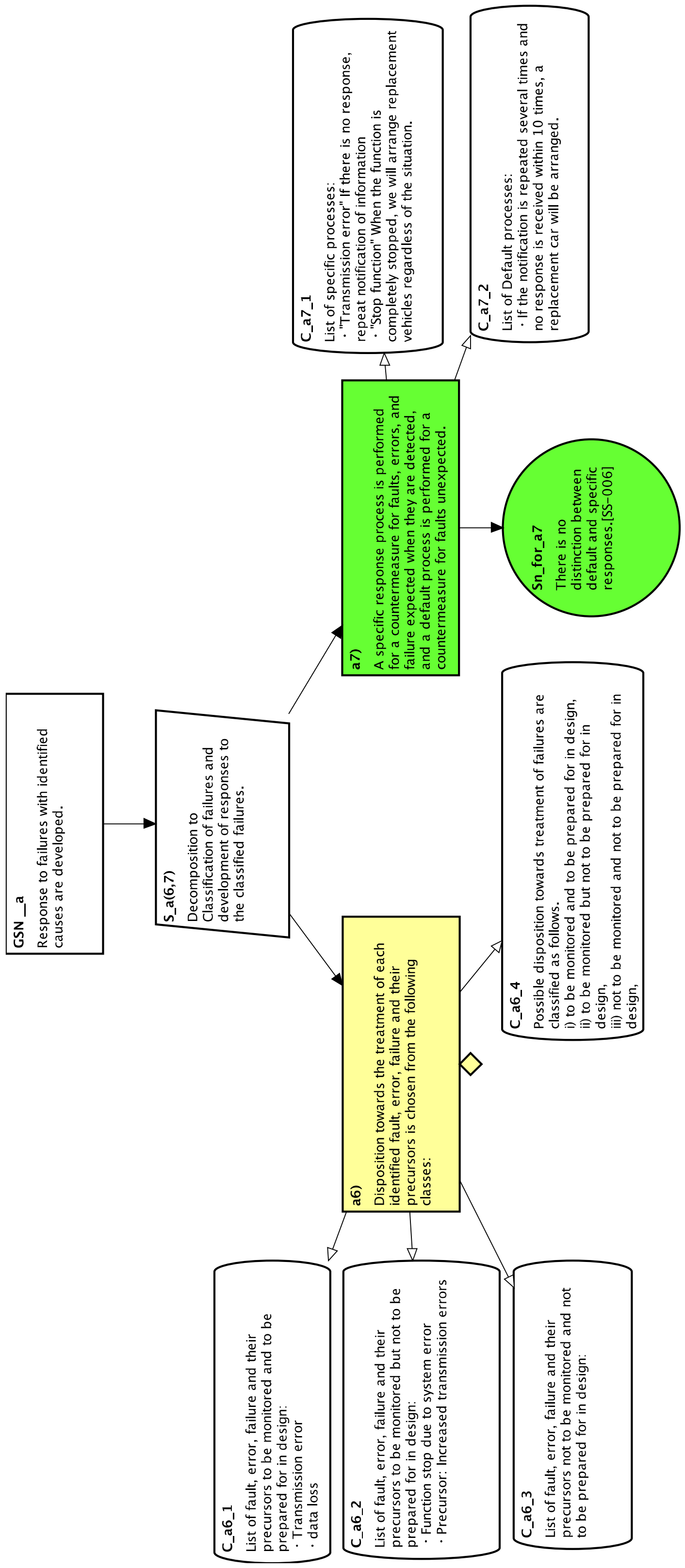

Figure 8: The detailed GSN for the outcome of GSN 'a) responses to failures with identified causes are developed.'

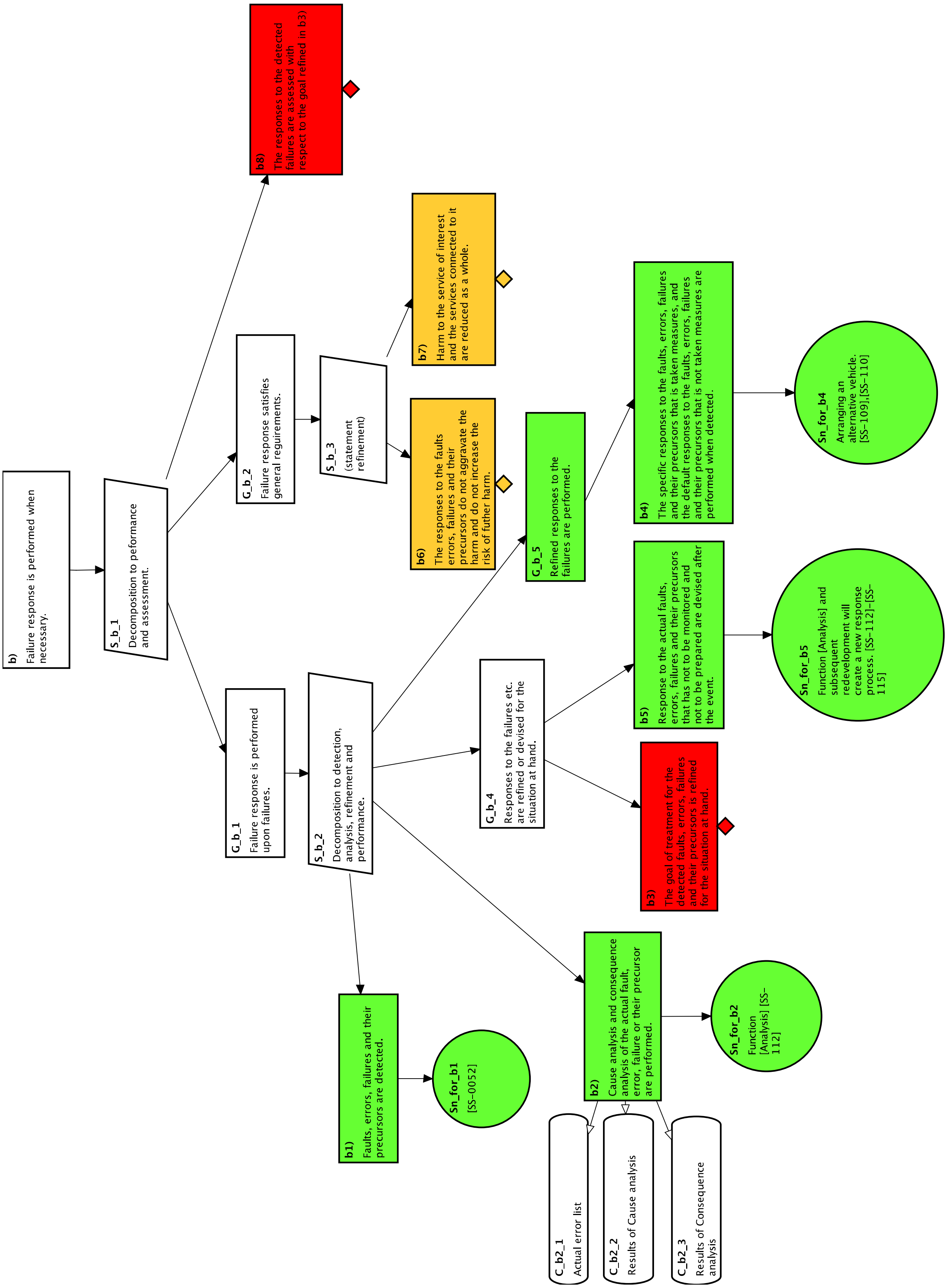

Figure 9: The GSN for the outcome of 'b) failure response is performed when necessary.'

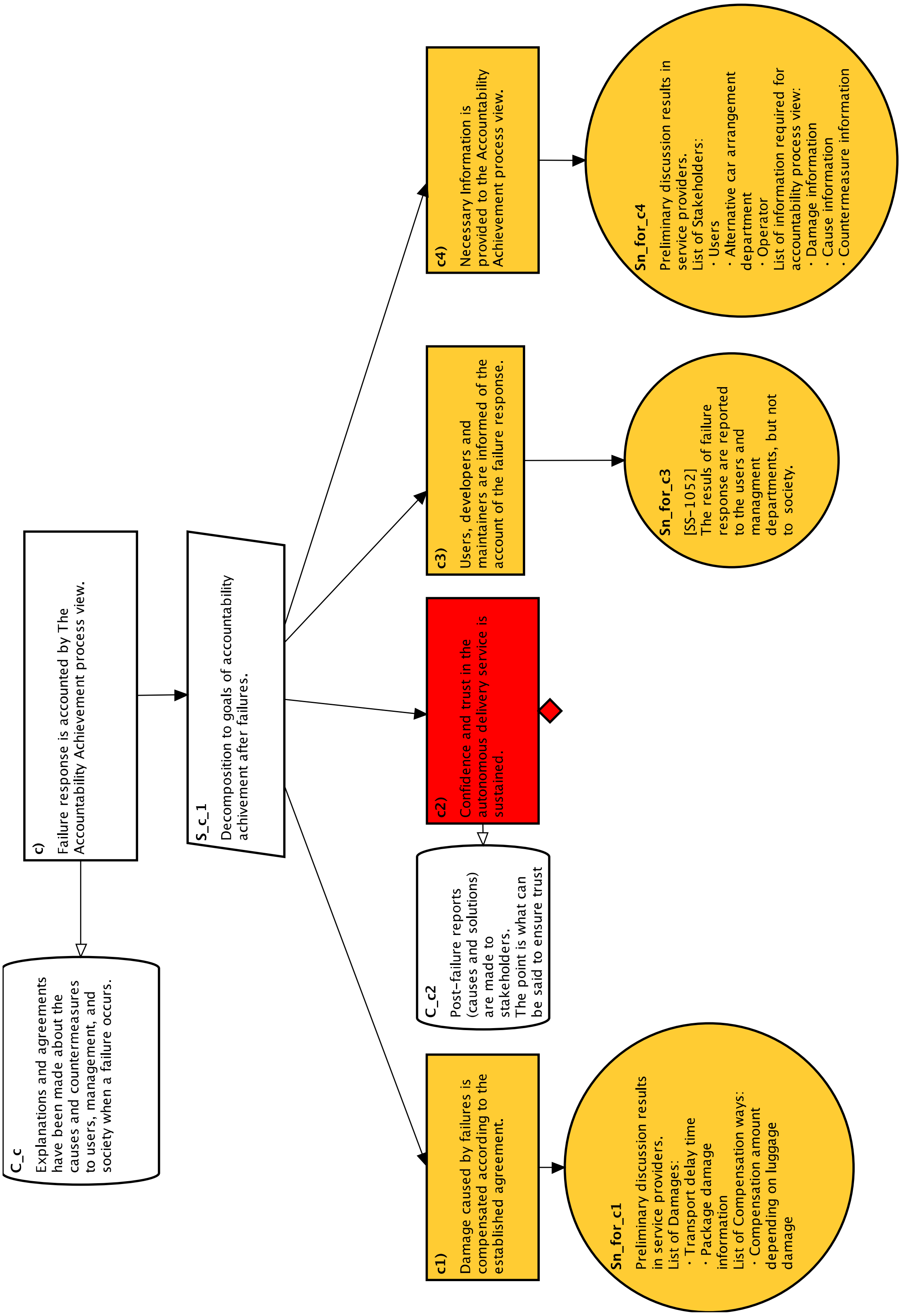

Figure 10: The GSN for the outcome of 'c) failure response is accounted by the accountability achievement process view.'

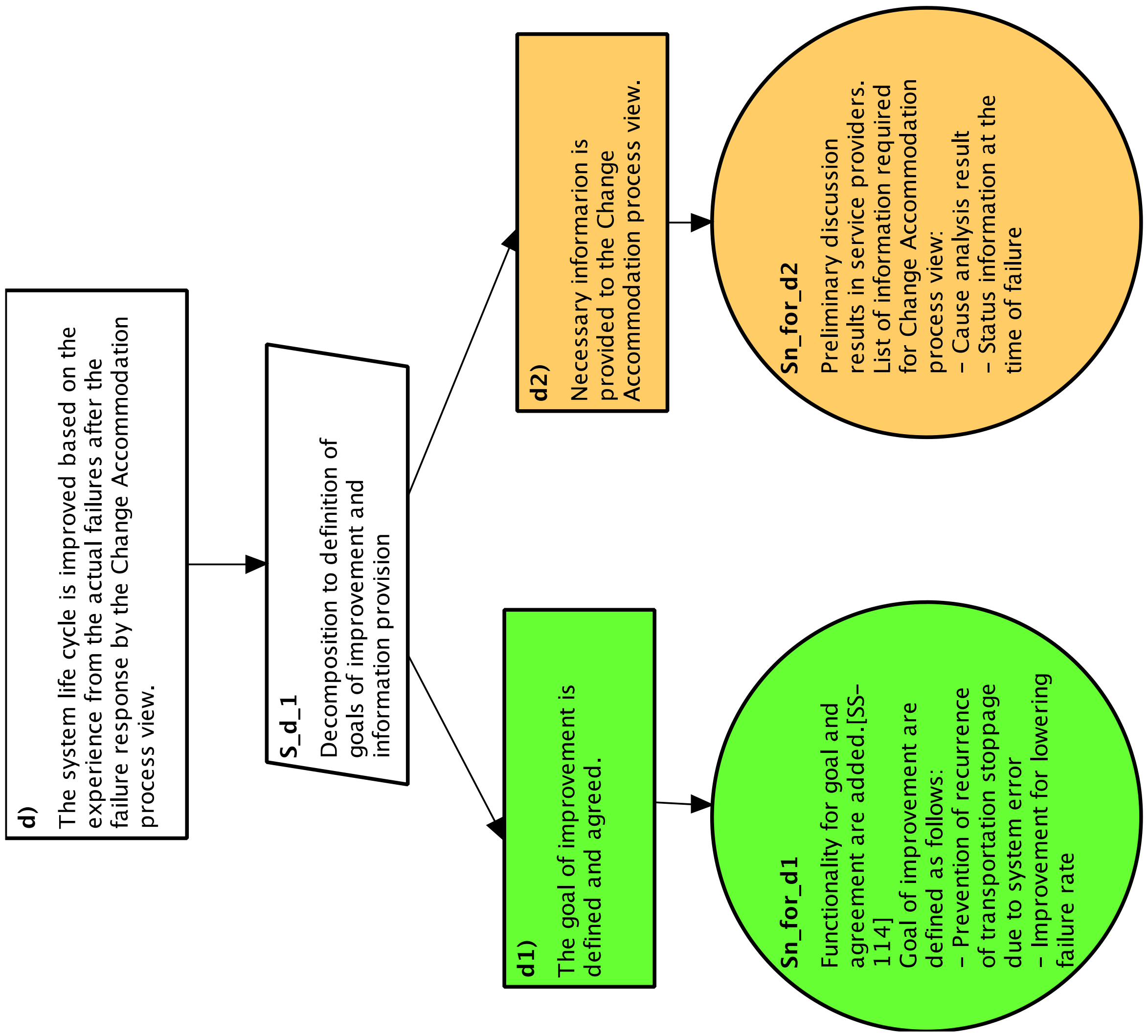

Figure 11: The GSN for the outcome of 'd) The system life cycle is improved based on the ... by the change accommodation process view.'

# TIGARS

Towards Identifying and closing Gaps in Assurance of autonomous Road vehicleS

**Project Ref: 01/18/05**

**Adelard Ref: W/3036/138008/28**

24 Waterside
44–48 Wharf Road
London
N1 7UX

T +44 20 7832 5850
F +44 20 7832 5870
E office@adelard.com
W www.adelard.com

Authors
Makoto Takeyama
Yoshiki Kinoshita

# TIGARS TOPIC NOTE 3: OPEN SYSTEMS PERSPECTIVE

## Summary

This TIGARS Topic Note provides a perspective on RASs assurance from the viewpoint of "open systems dependability" (OSD). We outline the issue for autonomous vehicles and provide some guidance on the deployment of the use of the open systems perspective from our work in TIGARS.

## Use of Document

The document is made available as a resource for the community, providing all use is adequately acknowledge and cited.

This document provides a snapshot of work in progress. We welcome feedback and interest in this work. Please contact the authors or admin.tigars@adelard.com

## Acknowledgement

This project is partially supported by the Assuring Autonomy International Programme, a partnership between Lloyd's Register Foundation and the University of York. Adelard acknowledges the additional support of the UK Department for Transport.

## Contents

## Figures

# 1 Introduction

This paper provides a perspective on RAS assurance from the viewpoint of "open systems dependability" (OSD) [1][2]. The OSD perspective shapes an approach, the OSD approach, to the assurance of RASs' trustworthiness in future, i.e., the assurance of claims of the type '*The deployed system will continue to perform as required in future*' (*OK in Future*). Trustworthiness in the future is an integral part of TIGARS's assurance framework presented in the TIGARS topic paper "Assurance - Overview and Issues (for D5.6)" [3]. The OSD approach influences all aspects of RAS assurance and system life cycles as any aspect at any time impacts RASs' future trustworthiness.

RASs possess features not covered in traditional approaches, making assuring "OK in future" difficult. Expectations and requirements on RASs are vague and keep changing along with changes in its environment. No single entity has full control or full understanding of an RAS. For example, the inner workings of ML components are typically opaque even to developers. An RAS's behaviour can change without its developers' involvement as it learns from its experience in the field and as, more mundanely, its ML components are updated by suppliers with effects unknown to the developers. For another example, how a deployed RAS interacts with other systems, physical environment, stakeholders and society in general cannot be delineated in advance and can change frequently in an unanticipated manner.

Thus, failures are inevitable. The possibility of glitches and failures of RASs cannot be eliminated even with the best traditional dependability efforts. Therefore, accountability in case of failures and continual improvement are essential for RAS.

The approach of open systems dependability is to establish in advance a regime, i.e., a system life cycle that is ready for eventual failures and that keeps improving through learning from experience.

# 2 Landscape – IEC 62853 Open systems dependability

IEC 62853:2018 Open systems dependability [1] is a recently introduced international standard providing requirements and guidelines for system life cycles of open systems to achieve dependability. It identifies and addresses four issues by means of four process views: consensus building, accountability achievement, failure response and change accommodation. A process view is a type of virtual life cycle process whose concept is described in ISO/IEC/IEEE 15288 [4] and 12207 [5].

The Consensus Building process view establishes and maintains explicit stakeholder agreements on the target RAS that prevent misunderstanding as much as possible. At the same time, it promotes more general common understanding among stakeholders that forms the basis for them to address eventualities unanticipated in the explicit agreements, which is inevitable for RASs.

The Accountability Achievement process view establishes in advance the relationship between a breach of an explicit agreement and its implication for stakeholders, which includes accountable stakeholders' obligation to provide remedies to non-accountable stakeholders. Assurance of accountability is crucial for acceptance of RAS deployment, which hinges on stakeholders' and the public's trust.

The Failure Response process view prepares for orderly responses to eventual failures of RASs. Performance of automatic responses planned at design-time and after-the-fact human intervention are integrated, as the former cannot be presumed perfect for RASs. The process view also ensures that operators' and developers' experience from failures leads to improvement of RASs, including recurrence prevention.

The Change Accommodation process view maintains the "fit for purpose" status of the target RAS despite changes in requirements, environments, objectives and/or purpose. These changes will be inevitable and frequent for the RAS as its standing in our sociotechnical world is far from established. It is crucial for the RAS to be able to address changes that cannot be anticipated in advance.

## 3 Applying open systems standards

Conceptual and abstract requirements in IEC 62853 need to be interpreted more concretely in the context of RASs. To that end, the following are being developed.

1. **DEOS Life Cycle Model**: a rigorous model of a reference system life cycle for achieving OSD;
2. **OSD Evidence Framework**: a framework for evidence documents for OSD assurance;
3. **OSD Deployment Platform**: an overall plan of implementation of the system life cycle

For Item 1, its mathematical formulation was described in [6] and its outline is given in Section 3.1. Items 2 and 3 are in planning stages as described in Section 3.2 and 3.3.

### 3.1 DEOS Life Cycle Model (DEOS-LCM)

The claim "OK in future" is about the life cycle of the system of interest, not about the system itself. However, assurance of life cycles is not well understood, due partly to the lack of consensus on what a "life cycle model" is. DEOS-LCM [6] is a model of system life cycle that provides the context of the assurance argument for the claim "OK in future".

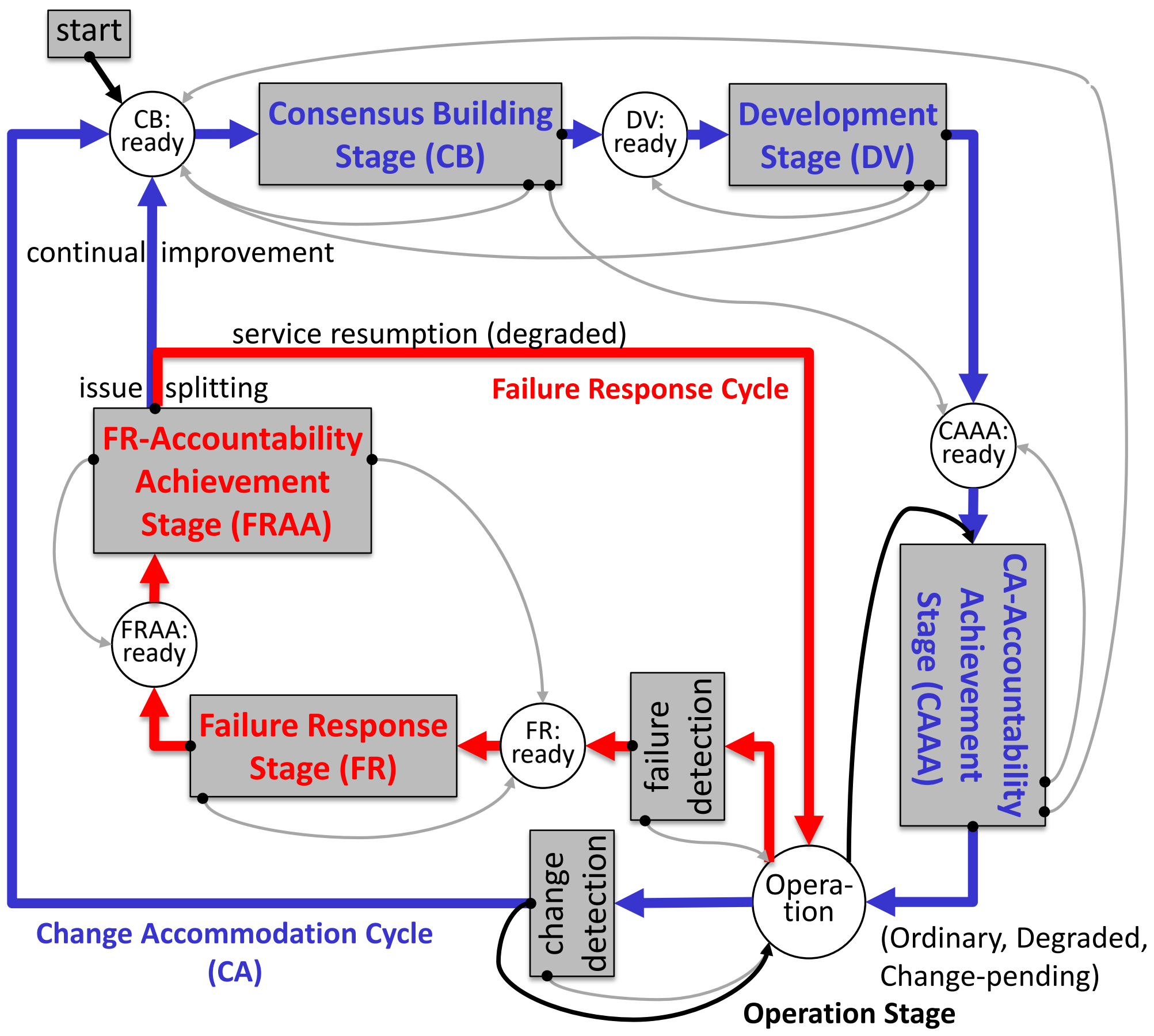

**Figure 1: DEOS Life Cycle Model (DEOS-LCM)**

The formulation of DEOS-LCM is based on the Dependent Petri Net (DPN). DPN is a variant of coloured Petri nets with inputs and outputs interacting and controlling the life cycle being modelled. Tokens are data representing *artefacts together with assurance* that they satisfy conditions associated with the place. Transitions between places correspond to life cycle (sub)stages that transform artefacts with assurance. Conditions associated with places are pre- and post-conditions of transitions. The "propositions as types" notion [7] is used to represent conditions and evidence (proofs) for assurance as data included in tokens. The Petri Net style is adopted to express necessary concurrent issues such as "promptly resume operation after a failure" and "develop an improved version to avoid failure recurrence, learning from experience".

DEOS-LCM is to be used as a workflow engine, taking as input artefacts produced, evidence that artefacts meet requirements, approval by authorities, etc. It checks the inputs and transitions to the appropriate next step, generating as outputs the results of checking, work-requests for the next step, etc. At any time, the target life cycle controlled by DEOS-LCM is achieving or will achieve the required outcomes of IEC 62853. The Assurance argument for what that is to be constructed can be assured on demand by using the assurance data in tokens.

Instantiating DEOS-LCM for the target system life cycle demands provides sufficiently precise identification and characterisation of artefacts, and other necessary information in the real world (including expert judgements and approvals of accountable stakeholders), so that rules for processing and decision making can be made explicit. The main purpose of DEOS-LCM formulation is to make it possible to examine those rules on a firm ground and to agree on them with least misinterpretation by all relevant stakeholders. Sophisticated automation for processing and decision making is not a motivation of DEOS-LCM.

## 3.2 OSD Evidence Framework (OSD-EF)

OSD-EF aims to identify and characterize assurance data for artefacts as demanded for in the instantiation of DEOS-LCM, such that the instance system life cycle is assured to achieve the outcomes of IEC 62853. OSD-EF is guidance on how to define, for artefacts at various life cycle stages, concrete requirements and evidence for them when adapting the OSD approach to a target system life cycle. It does so by extending the provisions for information items (documentations) in ISO/IEC/IEEE 15289 [8].

OSD-EF is based on the mappings between three established standards: IEC 62853, ISO/IEC/IEEE 15288, and ISO/IEC/IEEE 15289. The four process views of IEC 62853 guide how the OSD outcomes should be achieved using the processes of ISO/IEC/IEEE 15288. ISO/IEC/IEEE 15289 provides contents of information items consumed and produced by processes of ISO/IEC/IEEE 15288. The composition of the two relations relates the OSD outcomes with information items. Guidance in IEC 62853 on how a process should be performed translates to additional requirements on information items consumed or produced by the process. OSD-EF strengthens the contents of information items so that they include evidence for satisfaction of those additional requirements.

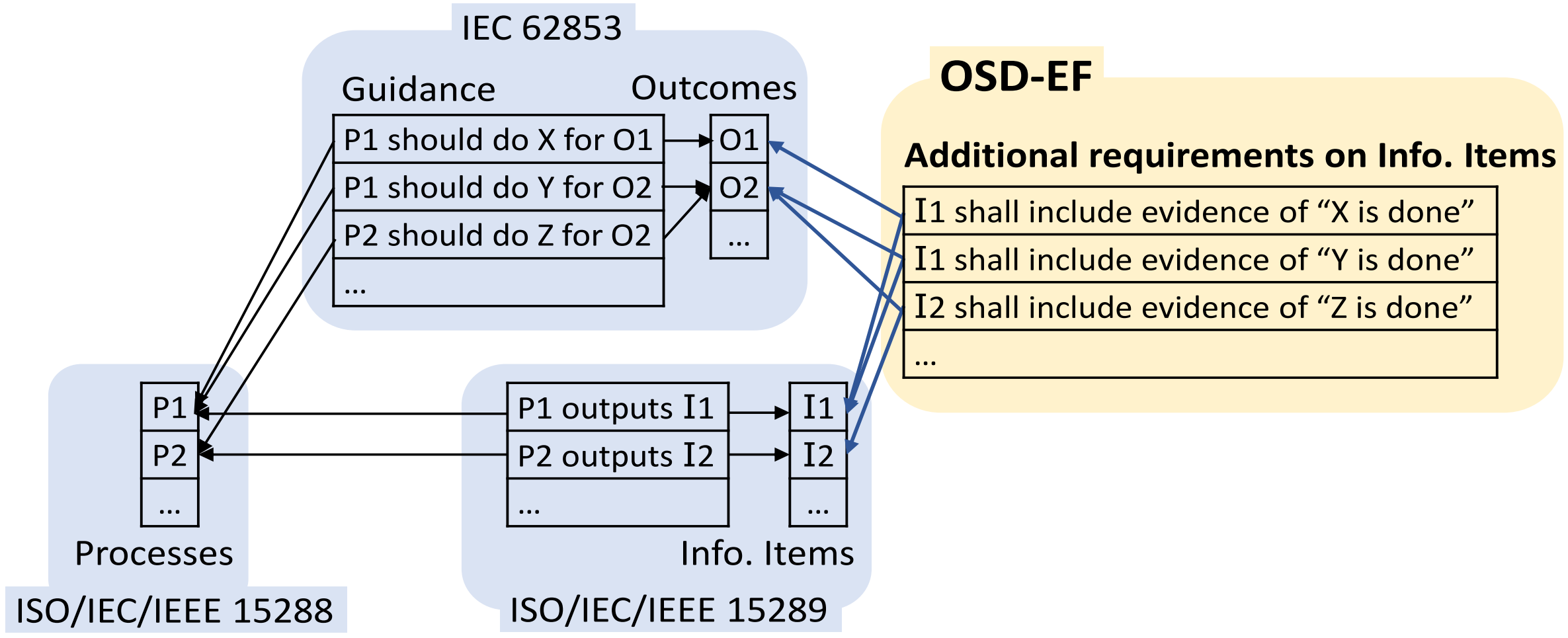

Figure 2: OSD Evidence Framework (OSD-EF)

## 3.3 OSD Deployment Platform (OSD-DP)

OSD Deployment Platform will be a common platform, i.e., a common overall architecture of development, operation and assurance of RASs that deploy DEOS-LCM and OSD-EF. OSD-DP will establish a consistent evaluation regime across RASs giving assurance that can be communicated, understood and trusted by a wide range of stakeholders and general public. OSD-DP will also enable assured operation of each RAS since it depends on assurance from RASs in the environment.

The general model of OSD-DP draws on those of security evaluation provided by ISO/IEC 15408 [9], of safety assurance by IEC 61508 [10], and of integrity levels by ISO/IEC 15026-3 [11]. OSD-DP introduces concepts of "Dependability Integrity Level" (DIL) and "Dependability Profile" (DP) as a basis of assurance. DILs, DPs, provisions of IEC 62853 and other related standards, OSD-EF, and DEOS-LCM are incorporated in the overall picture as shown below.

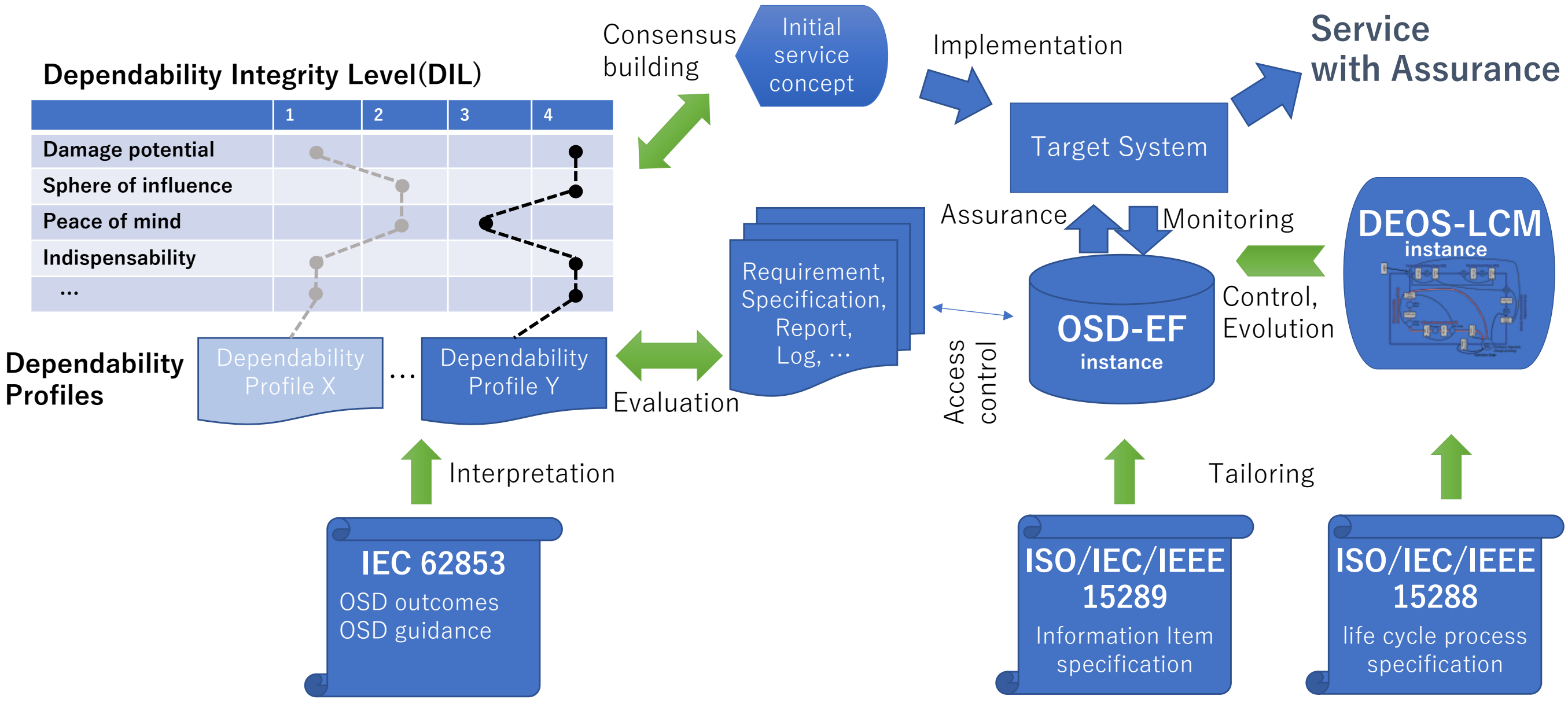

**Figure 3: OSD Deployment Platform (OSD-DP)**

A DIL represents a set of high-level dependability requirements and corresponds to an EAL (Evaluation Assurance Level) and a SIL (Safety Integrity Level). A DIL will be expressed as a simple tuple of component levels, e.g., (D4, S4, P3, I4), to aid communication, just as A-SILs are named A, B, C and D. Components of DILs are named such that a DIL conveys characteristics of target systems rather than requirements. A DP for a given DIL, which corresponds to a PP (Protection Profile), defines more concrete requirements, procedures, techniques, etc. that are necessary to achieve the DIL. Trade groups and such are expected to develop and agree on DILs' high-level requirements and on sector-standard DPs by interpreting and tailoring IEC 62853 in their context. DPs are expected to be validated and certified by appropriate authorities, similarly to PPs, and used as the basis of evaluation of individual systems.

For development of a new system, DPs provide a starting point for consensus building between stakeholders; namely, stakeholders negotiate and record consensus through selection / modification / new development of a DP. The life cycle of the system, the DEOS-LCM controlling it, and the OSD-EF are tailored to conform to the agreed DP. The OSD-EF data is stored online, linked to the requirements in DP and to the monitored status of the target system and environment. Based on that data, the DEOS-LCM controls operation and re-development processes of the target system, effects of which is reflected in the OSD-EF data. The OSD-EF data is access controlled according to the authorities of stakeholders who requires the data, enabling orderly communication. Thus, OSD-DP aims to maintain the state where evidence of conformance to the DP can be produced and evaluated continually, providing assurance at all times.

## 4 Guidance and recommendations for OSD perspective

We make the following recommendations:

1. RASs' trustworthiness in future should be assured systematically.

    - RASs are deployed in threat-filled environments that keep changing. For RASs to be accepted socially, stakeholders need to have confidence, before they are deployed, in how they are going to adapt to changes in future, including the response to and accountability for failures.

2. Projects on RASs should address the open systems dependability (OSD) of their RASs' system life cycles.

    - Assurance of RASs' trustworthiness in future must consider not only adaptation to future changes but also other factors affecting it. The OSD approach of IEC 62853 is one systematic approach to identifying and addressing four factors: consensus, accountability, failure response and change accommodation.

3. RAS projects should clarify and provide their accountability.

    - Accountability is a particularly important part of RAS assurance, since assurance of RASs' trustworthiness in future cannot be perfect and the accountability for potential failures, including provision for remedies, forms the major basis of RAS acceptance. Accountability is also a leading principle of the ethics of AI, as recognized in the latest recommendations of world organizations such as one [12] by OECD and another [13] by UNESCO.

4. RAS projects should take focused approaches to OSD to gain benefits with acceptable cost.

    - For example, one can focus on one level of system abstraction, such as the service level or the implementation level. TIGARS topic paper "Resilience and safety requirements" [14] focuses at the service level of a delivery RAS, successfully identifying and filling gaps between requirements and specifications. Another example is only one area of technologies being focused on, such as machine learning or communications.

5. Practical supporting material, such as examples, guidance and tools, for the application of IEC 62853 should be developed and provided to RAS projects, and also concrete lessons from experience on how to apply it should be accumulated.

    - Resulting best practices will lower the cost of achieving OSD for a specific RAS, which includes the cost of interpreting IEC 62853 in the RAS's context. Example interpretations, such as one for a delivery service RAS developed in the TIGARS topic paper [14], ease the effort for future RASs. Adopting and advancing the approach of DEOS-LCM [6], OSD-EF and OSD-DP will allow for more of a coordinated effort and systematic accumulation of knowledge.

6. Standardization bodies and industry trade groups should develop standards that augment and support IEC 62853.

    - One type of support standards could be to specialize IEC 62853 to particular industry sectors, such as automotive and assistive technology. Another type could standardize methods to realize OSD. DEOS-LCM [6], OSD-EF and OSD-DP are starting points of such standards.

7. Regulators should enable and promote the adoption of standardized OSD by RAS manufacturers, operators and users.

    - Once deployed, RASs of different origins interact with each other and form systems of systems without central controls. Adverse consequences are possible even if each RAS manufacturer or operator has no malicious intent. Realizing that OSD is the very means to counter such consequences, but this depends on the ability of RASs to collaborate with each other. Regulators play a key role in ensuring such collaboration is possible across legitimate RASs. Sharing of

information, including assurance, is a major concern. Regulators can incentivize RAS manufactures and operators to participate in an evaluation and assurance regime, such as the one that is proposed by OSD-DP, which in turn could strengthens users' confidence in RASs' trustworthiness in future.

8. The concept of OSD itself should evolve, reflecting findings from applications to RASs.

    - To address security, the Change Accommodation process view itself should be able to change and adapt to new threats. It was found through the development of the generic RAS assurance case [3] that combines the OSD approach and the security informed safety of PAS 11281 [15] that the Accountability Achievement process view should balance transparency of information against security, intellectual property protection, privacy, etc., when it is applied to regimes of type approval and inspection that cover autonomous vehicles such as Road Transport Vehicle Act of Japan. The Consensus Building process view should help manage the implications of trade-offs, which are often at the core of problems.

# 5 Bibliography

[1] IEC. IEC 62853:2018 *Open systems dependability*, 2018.

[2] M. Tokoro, ed. Open systems dependability: dependability engineering for ever-changing systems. CRC press, 2015.

[3] TIGARS project. Assurance – Overview and Issues (for D5.6). *TIGARS Synthesis of Project Results*, D5.6.1, 2019.

[4] ISO, IEC, and IEEE. ISO/IEC/IEEE 15288:2015 Systems and software engineering — System life cycle processes, 2015.

[5] ISO, IEC, and IEEE. ISO/IEC/IEEE 12207:2017 Systems and software engineering — Software life cycle processes, 2017.

[6] S. Kinoshita, Y. Kinoshita, and M. Takeyama. A modelling approach for system life cycles assurance. In *International Conference on Computer Safety, Reliability, and Security*, pages 16–27. Springer, 2019.

[7] B. Nordström, K. Petersson, and J. M. Smith. *Programming in Martin-Löf's type theory*, volume 200. Oxford University Press Oxford, 1990.

[8] ISO, IEC, and IEEE. ISO/IEC/IEEE 15289:2019 Systems and software engineering — Content of life cycle information items (documentation), 2019.

[9] ISO and IEC. ISO/IEC 15408 Information technology — Security techniques — Evaluation criteria for IT security, 2009.

[10] IEC. IEC 61508 Functional safety of electrical/electronic/programmable electronic safety-related systems, 2010.

[11] ISO and IEC. ISO/IEC 15026-3 Systems and software engineering — Systems and software assurance — Part 3: System integrity levels, 2015.

[12] OECD (Organisation for Economic Co-operation and Development). *Recommendation on Artificial Intelligence (AI)*, OECD/LEGAL/0449. *OECD Legal Instruments*, 22 May 2019. Available from: https://legalinstruments.oecd.org/en/instruments/OECD-LEGAL-0449

[13] UNESCO. Preliminary study on a possible standard-setting instrument on the ethics of artificial intelligence. *UNESCO General Conference, 40th Session*, 40 C/67, 30 Jul 2019. Available from: https://unesdoc.unesco.org/ark:/48223/pf0000369455?posInSet=2&queryId=d507ea71-98ec-444b-ae93-06ad1482f339

[14] TIGARS project. Resilience and safety requirements. *TIGARS Synthesis of Project Results*, D5.6.2, 2019.

[15] BSI. PAS 11281:2018 *Connected automotive ecosystems — Impact of security on safety — Code of practice*, 2018. Available from: https://shop.bsigroup.com/ProductDetail?pid=000000000030365540

# TIGARS

Towards Identifying and closing Gaps in Assurance of autonomous Road vehicleS

**Project Ref: 01/18/05**

**Adelard Ref: W/3014/138008/20**

# TIGARS TOPIC NOTE 4: FORMAL VERIFICATION AND STATIC ANALYSIS OF ML SYSTEMS

## Summary

In this paper, we scope and assess the existing gaps and challenges of deploying state-of-the-art static analysis and formal verification techniques to ML models that may be deployed in autonomous vehicles. We present preliminary results regarding the applicability of state-of-the-art formal verification and static analysis to ML algorithms and supporting systems, respectively.

24 Waterside
44–48 Wharf Road
London
N1 7UX

T +44 20 7832 5850
F +44 20 7832 5870
E office@adelard.com
W www.adelard.com

Authors
Heidy Khlaaf
Philippa Ryan

## Use of Document

The document is made available as a resource for the community providing all use is adequately acknowledge and cited.

This document provides a snapshot of work in progress. We welcome feedback and interest in this work. Please contact the authors or admin.tigars@adelard.com

## Acknowledgement

This project is partially supported by the Assuring Autonomy International Programme, a partnership between Lloyd's Register Foundation and the University of York. Adelard acknowledges the additional support of the UK Department for Transport.

## Contents

## Figures

# 1 Introduction

Formal verification and static analysis techniques have been widely deployed on traditional safety-critical systems for several decades as part of verification & validation (V&V) processes. However, the rapid introduction of Machine Learning (ML) systems in these environments poses a great challenge from both a regulatory and system assurance view. The lack of applicable verification & validation techniques for ML systems stifles existing assurance strategies, curbing the potential for innovation and benefits to be gained from their deployment.

Although a flurry of preliminary research techniques to verify ML algorithms have recently permeated academic literature, it has been unclear as to whether and how these novel methods contribute to the disparate sources of evidence needed in assurance cases, in particular, for safety-critical systems such as autonomous vehicles. In this paper, we thus aim to assess V&V gaps of the existing methods, and focus on the need to support the assurance claim: that the required behaviour and functionality of the system are defined and valid, and that the system behaves according to its requirements when deployed. Although extensive literature exists regarding more general V&V techniques for complex cyber-physical systems [20], this is beyond the scope of this paper, as we specifically focus on the use of ML algorithms within these systems.

We focus on directly investigating the desired behaviour (e.g., the safety property or reliability) of a system through an outcome-based approach. There are a variety of ways in which the desired properties of a system can be classified. Indeed, there has been considerable work defining the dependability terminology (e.g., [1]). The chosen catalogue of behavioural attributes currently used in the assessment of safety relevant components, and what we will consider in the remaining sections, are: functionality, performance, reliability, operability, robustness, availability, and security.

These attributes may overlap individually, and also depending on the application at hand. In the context of autonomous systems that utilize ML algorithms, the overlap between security and dependability (reliability, operability, robustness, and availability) is extensive. As demonstrated in [2], Confidentiality, Integrity, and Availability (CIA) attributes are closely tied to and affect the functionality, performance, and the overall dependability of the system.

In the remaining sections, we address some of the challenges faced in the current landscape regarding the V&V of the noted attributes for ML systems. Additionally, we discuss potential solutions and recommendations proposed by a varied set of literature as well as preliminary research that we have carried out.

# 2 Formal verification of ML models

Formal verification is the process of establishing whether a system satisfies some requirements (properties), using formal methods of mathematics. Traditionally, formal methods are based on the premise that we can determine the functional properties of a system by the way we design it and implement it. While this holds for traditional systems, it does not hold for ML systems as a whole, as their design determines how they learn, but not what they will learn [11]. Furthermore, formal verification methods are based on the premise that we can infer functional properties of a software product from an analysis of its source code. While this holds for traditional systems, it does not hold for ML systems, whose behaviour is also determined by their learning theory. Indeed, it is the case that traditional formal methods techniques cannot be applied as they are. Novel verification techniques and specifications thus must be devised to address ML algorithms, specifically, for Neural Networks (NNs).

## 2.1 Novel vulnerability and attack avenues

Besides the difficulty of applying existing specifications and verification techniques to ML algorithms, ML systems pose new attack surfaces and threats in which adversaries aim to: influence and exploit the collection and processing of data, corrupt the model, and manipulate the resulting outputs [2]. Notoriously, researchers in [3] have forced models to make wrong predictions by computing what are now known as *adversarial examples*. These are examples that produce perturbations that are very slight and often indistinguishable to humans, yet are sufficient to change the model's prediction to one that is incorrect. Unfortunately, these perturbations can now be efficiently and rapidly produced through the fast gradient sign method introduced in [4], as shown in Figure 1.

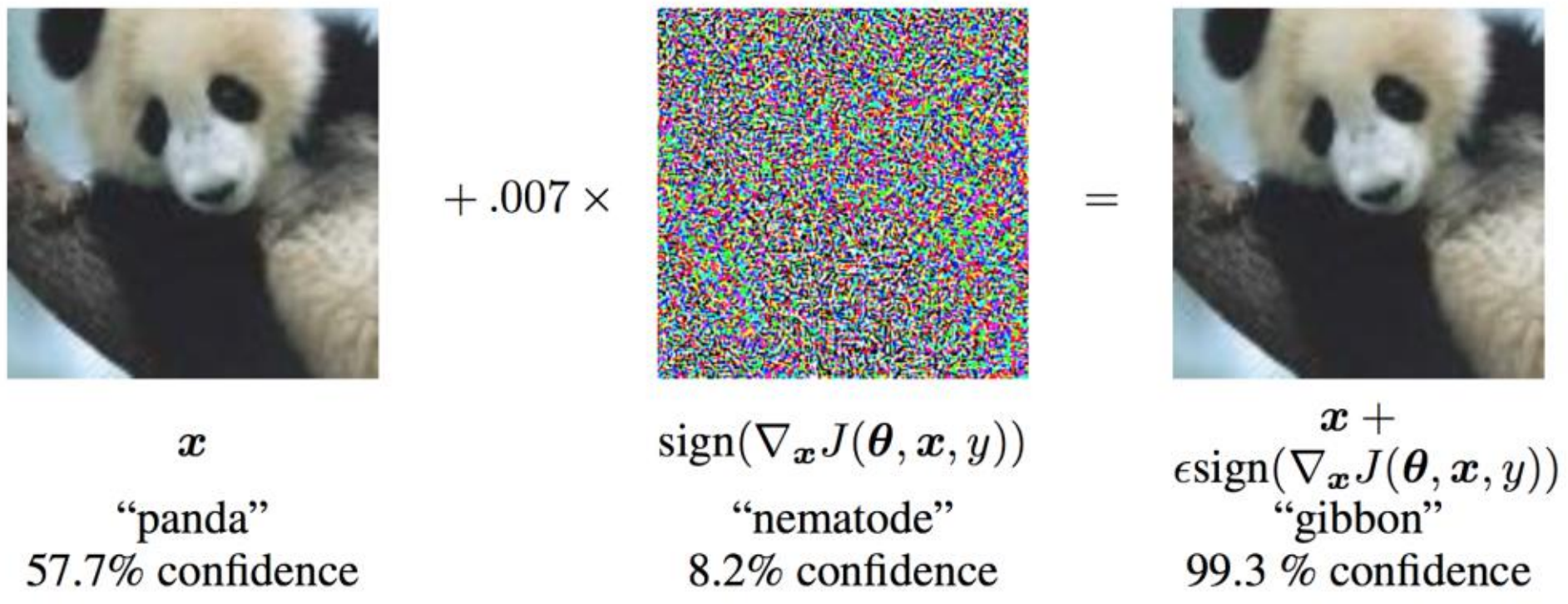

**Figure 1: The fast gradient sign method introduced by [4].**

Adversaries are capable of manipulating the model inputs to affect its output, thus reducing the robustness, accuracy, availability, and integrity of the overall behaviour of the system. This is due to ML models not being robust against input distribution drifts, where the training and test distributions differ. We discuss these novel attack avenues and preliminary potential solutions.

## 2.2 Current landscape

Research to create formal methods that can verify ML models is in its infancy, due to the lack of formal specifications that can address the attributes discussed in Section 1 for ML algorithms and corresponding verification techniques. The difficulty arises given that the output behaviour of an ML algorithm may not always be clear or expected relative to the inputs. To consider the novel complexities of ML systems, traditional formal properties thus must be reconceived and redeveloped for ML models.

Some preliminary research has attempted to find ways of specifying types of ML robustness (i.e., pointwise robustness) that would be amenable to formal verification, which aims to verify an ML model's robustness against the adversarial examples noted in Section 2.3. However, the remaining system dependability properties noted in Section 1 have gone unspecified. Furthermore, recent methodologies have not accounted for specifications unique to ML-based systems such as accountability, fairness, and privacy. It has been shown that sensitive personal data, if used in training, can be extracted from the ML model outputs [5][6]. In a way, ML models capture and encompass elements of their training data, thus it is no surprise that their confidentiality and privacy are at risk.

The majority of researchers have thus concentrated their efforts on the verification of *pointwise robustness, given its availability as the only ML oriented specification*. Below, we define pointwise robustness further and describe the preliminary set of state-of-the-art research available to verify it. Subsequently, we discuss the shortcomings of the property and its verification techniques, and outline other methodologies and the next steps that need to be taken to mitigate for the gaps identified in formal verification research.

## 2.3 Pointwise robustness

*Pointwise robustness* aims to identify how sensitive a classifier function is against input distribution drifts (i.e., small perturbations as shown in Figure 1). Formally, it is the property in which a classifier function $f'$ is not robust at point $x$ if there exists a point $y$ within mathematically defined distance $\eta$ such that the classification of $y$ is not the same as the classification of $x$. That is, for some point $x$ from the input, the classification label remains constant within the "neighbourhood" $\eta$ of $x$, even when small value deltas (i.e., perturbations) are applied to $x$. A point $x$ would not be robust if it is at a decision boundary and adding a perturbation would cause it to be classified in the next class. Generally speaking, the idea is that a "neighbourhood" $\eta$ should be reasonably classified as the given class.

Despite being the only ML oriented specification available, the verification of a sub-property such as pointwise robustness has proven to be difficult. Verification techniques introduced thus far are often specific to the ML model at hand, require manual reasoning, and are not scalable. We discuss the main contributions introduced in the field thus far.

Pulina et al. developed the first verification technique for demonstrating robustness [7], in which the output class of a neural network is constant across a desired neighbourhood. However, this technique is limited to only one hidden layer in a Multi-Layer Perceptrons (MLPs) network. Furthermore, their chosen case study was a network with less than a dozen hidden neurons. Their strategy relied on over-approximating the sigmoid activation function using constraints to reduce the problem to a Boolean satisfiability problem.

Huang et al. built on the above initial method and proposed a new verification method applicable to deep neural networks and other neural networks [8]. This technique is more scalable, as it accounted for six layers with hundreds of neurons in a case study. However, this technique relies on the assumption that only a subset of the hidden neurons in the neural network are relevant to each input. An adversary (especially a strong one) can violate this assumption by manipulating one of the neurons that was assumed to be irrelevant to evade detection.

Finally, and most notably, the tool Reluplex [9] focuses on rectified linear networks, and exploits their piecewise linear structure to produce constraints to be fed into a specialized linear programming solver. Reluplex is not as scalable as the technique introduced in Huang et al., but the Reluplex authors claim to be able to verify functional properties as well as pointwise robustness (see discussion in Section 2.4).

No matter the technique, these verification methods suffer from the same set of limitations:

- it is difficult to define meaningful regions ($\eta$) and manipulations
- the neighbourhoods surrounding a point ($x$) that are used currently are arbitrary and conservative
- we cannot enumerate all points near which the classifier should be approximately constant, that is, we cannot predict all future inputs

Some preliminary work extending [8] has been introduced, proposing that instead of relying on an exhaustive search of a discretized region (i.e., $\eta$), one can compute the upper and lower bound case confidence values of a point $x$ [10], which may alleviate some of these limitations.

### 2.3.1 Drawbacks of pointwise robustness

Generally speaking, pointwise robustness, although an interesting property, is not expressive enough nor conducive to producing confidence for assuring an ML model, given that one cannot predict all future inputs. Furthermore, pointwise robustness solely focuses on indistinguishable perturbed inputs, and thus implicitly assumes a niche attack model in which an attacker is given an input image from a data distribution, in which he or she must perturb said image in a way that is undetectable by humans. Indeed, the authors of the fast gradient sign method, introduced in [4], were unable to find a compelling example that required indistinguishability [13] within the security attack model introduced in [2]. They outline the standard rules assumed in the perturbation defence literature, and detail why pointwise robustness is an insufficient measure of both security and robustness.

Recall that pointwise robustness aims to verify that for some point $x$ from the input, the classification label remains constant within the "neighbourhood" $\eta$ of $x$, even when small value deltas (i.e., perturbations) are applied to $x$. In the ML literature, these perturbations are often known as $l_p$ perturbations, in reference to Lebesgue spaces. Indeed, sufficiently small $l_p$ perturbations will produce examples that are indistinguishable to humans. However, research has shown that $l_p$ is a poor proxy for measuring what humans actually see [14]. It is also the case that larger $l_p$ norms could produce indistinguishable perturbations, whether or not they pose a real threat to robustness or security. Modifying the neighbourhood size $\eta$, relative to $x$, does not provide any solutions, as the images exactly size $\eta$ away from an input image will include perceptually different images, as well as images that may take a long time to detect as different, as shown in Figure 2 below. Given that $l_p$ is a poor proxy for measuring what humans see, $\eta$ thus cannot be an indicative measure of such a discrepancy.

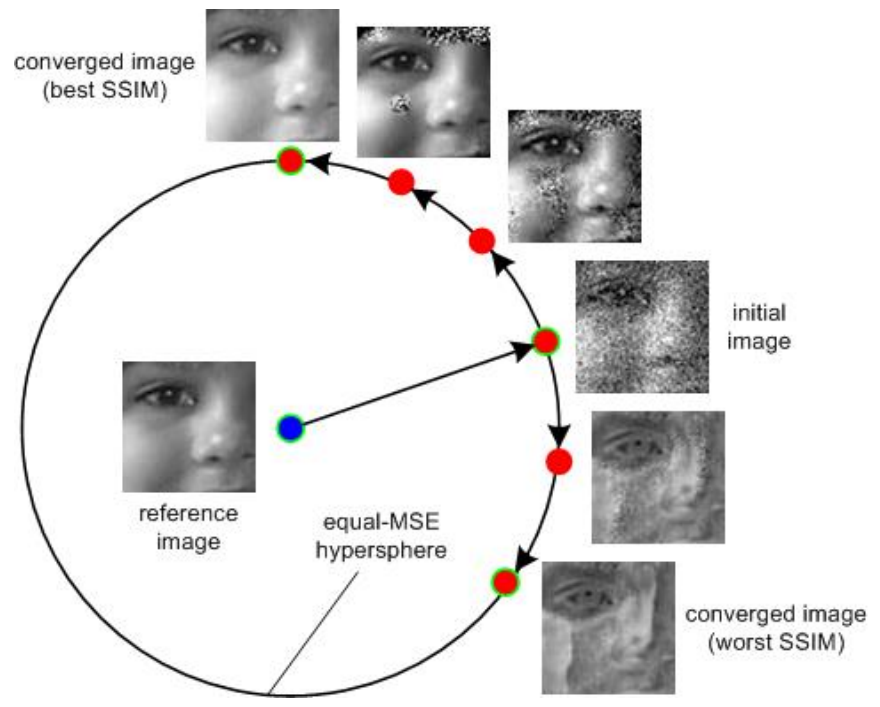

**Figure 2: Images equally far away from an input image within the same neighbourhood can drastically differ [14]**

Furthermore, [19] demonstrates that adversarial examples with a small Hamming distance ($l_0$) are a natural by-product whenever we partition the high dimensional input space into a bounded number of labelled regions by ReLU neural networks.

Finally, works discussing adversarial perturbation and pointwise robustness not only fail to note real-world examples, but also how state-of-the-art verification techniques can be applied to real-time systems. Additionally, the same outcomes derived from adversarial indistinguishable perturbations can be achieved by much simpler attacks that do not require machine learning components [13] (e.g., simply covering or physically perturbing a stop sign, in the case of autonomous systems). It is thus unclear what assurances, if any, a pointwise robustness analysis would provide, given its lack of realistic adversarial action spaces. With regard to the safety assurance of an autonomous system, we fail to see the claim for which pointwise robustness can provide support or evidence.

## 2.4 Formally verifying well-specified systems

Unlike many of the techniques introduced in Section 2.3, Reluplex [9] is an outlier in that its authors aim to verify more general behaviours regarding ML algorithms, instead of just pointwise robustness. The Reluplex technique requires functional specifications, written as constraints, to be fed into a specialized linear programming solver to be verified against a piecewise linear constraint model of the ML algorithm. The generalization of this algorithm is challenging, as it requires an existing well-defined, bounded, non-ML system to be in place, or at least specified. The Reluplex case study is carried out against an unmanned variant of ACAS X, known as ACAS Xu, which produces horizontal manoeuvre advisories. The ACAS Xu has already been implemented (and thus specified) using traditional software (e.g., non-ML) components utilizing a large lookup table that maps sensor measurements to advisories. The specification constraints defined in [9] are limited to the functional requirements of the ACAS Xu, that is, they are traditional system requirements, devoid of specifications unique to ML algorithms (recent improvements and expansion to the tool can be found in [29]).

The Reluplex technique provides a promising solution for well-specified traditional systems requiring new ML implementations. However, related work already existed in the safety-critical domain: specifically, the numerous deterministic methods for the verification of neural nets trained to approximate look-up mappings in civil and space aviation [11]. We anticipate that some implementations of the planning unit in an autonomous system may be within the use bounds of the Reluplex technique. However, the generalization of Reluplex and similar techniques to deep learning is challenging because they require well-defined, mathematically specifiable system specifications as input. These techniques are thus only applicable to well-specified deterministic or tractable systems that can be implemented using traditional methods (e.g., programming in C) or via ML models.

As a consequence, these techniques cannot be straightforwardly applied to arbitrary contexts, and domain-specific effort is currently required even to specify properties of interest, let alone verify them. Consider for example perception systems (e.g., camera, LIDARs, sensor fusion) crucial to the operation of autonomous vehicles: their system specifications are not bounded, and are not amenable to formalization of constraints that can be proved.

# 3 Static analysis of supporting systems

Autonomous systems, including autonomous vehicles, contain more than just the ML components. Traditional systems exist throughout, either separate from the ML functionality or in support of it. Typical vulnerabilities faced by the non-ML supporting systems must still be considered against all the attributes under consideration, as supporting software is paramount to building ML models, whether it be through using pre-existing or in-house libraries to manage and implement a neural network. Static analysis is the term used for source and object code analysis that examines the code without execution, and may encompass some formal verification techniques. These analyses do semantic and syntactic checks on the source and object code.

Static analysis can be performed on supporting software to look for potential issues and vulnerabilities that could impact the performance or functionality of the code. In this section, we examine how static analysis can be adapted for this area. We then present a preliminary analysis of some open source ML vision software to demonstrate the applicability of these methods.

## 3.1 Vulnerability analysis

In general, the threats faced by the supporting systems differ greatly to those faced by ML systems, and extensive literature and research in the past decades have addressed a myriad of these issues. A review of the existing threats to traditional systems is beyond the scope of this deliverable. However, researchers have found that ML software is indeed vulnerable to traditional threats [12]. Furthermore, researchers have shown that overflow errors within supporting software can propagate and affect the functionality of an ML model, as they identified an issue in a robotics system where a Not a Number (NaN) code error could cause uncontrolled acceleration [15].

Run-time issues such as overflow/underflow and access to data out of bounds can directly impact on the performance of an ML element such as a NN, as it may perform a series of matrix multiplications on edges and nodes. That is, for ML it may be more difficult to identify which parts of the software are affected by the error, and hence what the impact might be. Consider a loop that can potentially access data outside an array. For a traditionally developed system, the final purpose of that array and the loop are likely to be relatively transparent (even when found in a generic library function). However, for ML, the use of undefined data in one array manipulation may or may not have an impact, depending on the sensitivity to that data point. Additionally, overflow/underflow in an NN could lead to edges or nodes having a multiplication factor of zero for some operations. Again, it is difficult to predict the impact this might have on a task such as classification except in a very broad sense.

Note that the aforementioned errors would only be applicable to statically-typed languages such as C and C++. A case study is thus carried out in Section 3.2 to further analyse the implications and impact of such errors on an open source ML library implemented in C. In related work, other researchers have demonstrated a methodology in which developers can use an interactive proof assistant to implement their ML system and prove formal theorems that their implementation is devoid of errors [21].

We note that the aforementioned errors would not be applicable to a language such Python, a popular language utilized in the implementation of numerous ML libraries and frameworks, as the semantics and implementation of the language itself prevent overflow/underflow errors. However, Python is a dynamically typed language, bringing about a different set of program errors not exhibited by statically typed languages such as C and C++. We discuss this issue further in Section 3.3.

## 3.2 Preliminary case study

In order to explore the applicability of static analysis to ML subsystems and supporting infrastructure, we have performed preliminary static analysis experiments on the You Only Look Once (YOLO) [16] CNN source code using Polyspace Bug Finder [17]. Polyspace is an industry standard tool, used by many developers of high integrity software, to look for potential bugs and to assure compliance with the MISRA coding guidelines [31]. We selected YOLO in our case study, as its source code is publicly available, and it was used by Witz and Nagoya in an early prototype of the golf buggy speed control system [30].

The analysis identified potential modifications that could be made to the off-the-shelf (OTS) software, improving robustness without impacting on the deployed functionality. This could support a case for high integrity use, along with additional evidence of the appropriateness of the training. Note that security concerns of using a public source image training set should also be considered, as this could provide an attack vector. However, our experiments were limited to examining the underlying source code only.

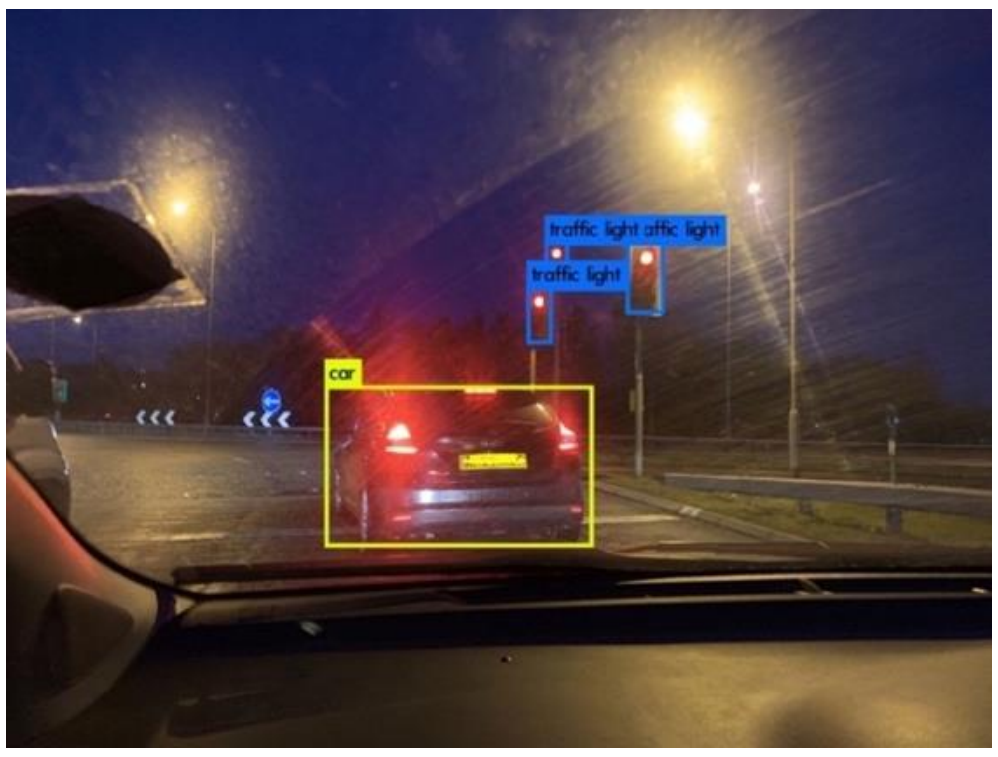

**Figure 3: Example object recognition support by YOLO on public road**

We analysed the core C and C++ software YOLO software library, including

- the main classification application, which loads a weights and configuration file (in this case trained using COCO) and applies it to a specified image file
- YOLO libraries including training code (which would not typically be used during operation but would impact the robustness/reliability of the output), image manipulations and matrix calculations
- third-party GPU source code from NVIDIA as the library supports multi-threaded training using a GPU
- third-party code for accessing a webcam for "live streaming" of input/output

The latter two items were included so that the potential impact of concurrency problems could be considered.

Run time errors results

A number of different run-time errors were identified. Issues that could be of concern included:

- Known security vulnerabilities such as file I/O, which could be hijacked for system files, and buffer/data vulnerabilities allowing memory to be overwritten.
- A number of memory leaks, such as files opened and not closed, and temporarily allocated data not freed, which could lead to unpredictable behaviour, crashes, and corrupted data.
- A large number of calls to `free()` where the validity of its use is not checked. This could lead to incorrect (but potentially plausible) weights being loaded to the network.
- Potential "divide by zeros" in the training code, including cost calculation. This could lead to crashes during online training, if the system were to be used in such a way.

MISRA 2012 analysis

Compliance to the mandatory MISRA 2012 guidelines was assessed. Violation of these rules often means that bugs are present in the code. A summary is presented below.

- Rule 21.17 – violations that could lead to undefined behaviour if the input configuration files were corrupted or had invalid data. The impact of this might be undefined behaviour, either causing YOLO to crash or not be started.
- Rule 21.18 – violations that could lead to memory leaks. Again, this is a potential reliability issue.

Mitigations would be to have systems using YOLO detecting its liveness and having a safe state response if possible.

Concurrency analysis results

Polyspace identified a potentially conflicting access to a buffered image array during concurrent display/processing of a live webcam stream. The code is in a demo of functionality but for the purposes of exploration we have assumed it could be used, for example, to display to a car operator the output of the classifier. This could have two issues:

1. If the display is being used by an operator and the data appears in a corrupted form it could reduce confidence in the image classification and the operator would take control (if this option was available).
2. If the data being used during image classification was corrupted then items could be missed/mis-classified/detected and classified when they did not exist. This is potentially more serious, but may depend on repeated identical failures input into a decision-making element. The decision-making element would need awareness of this type of failure mode and to have an appropriate mitigating action.

Running the YOLO source code through Polyspace indicated a number of potential issues that could impact the reliability and also the safety of the code if it were to be implemented as part of a safety critical system. The case study has demonstrated that traditional approaches to static analysis are still applicable to the supporting software, and could help make the code more trustworthy in operation and improve an assurance case for its use. However, further analysis is still needed, for example to consider the impact of mathematical bugs in training routines which might have impacted the CNN functionality derived. Our next step in the analysis will be to perform a structured walkthrough of the code to identify which key functions during classification and training are impacted by the identified bugs.

## 3.3 Static analysis of dynamically typed languages

Although we have discussed the potential use of existing static analysis methods, unfortunately, the majority of these techniques and analyses apply to statically typed languages such as C and C++. However, Python, a dynamically typed language, has unexpectedly infiltrated the safety-critical domain given the recent use of Python within ML frameworks (e.g., TensorFlow, PyTorch, etc.). These frameworks can be deployed in safety-critical contexts, such as autonomous vehicles, yet static analysis and formal verification techniques for the Python language are almost non-existent. The lack of existing techniques is due to a few factors:

- Dynamically and weakly typed languages are often discouraged from use in safety-critical domains given common type faults they may produce.
- Python's lack of use in critical domains has never incentivised the creation of novel formal analysis techniques addressing dynamically typed languages.
- Dynamically typed languages are significantly harder to statically analyse, or verify, given the difficulty of constructing control or data flow graphs, which are easier to derive from languages such as C/C++.

Nonetheless, the use of Python in the safety domain is now prevalent, given that it is the language of choice for numerous ML libraries. However, formal methods techniques (e.g., static analysis, or formal verification) have proved to be challenging to produce for dynamically typed languages.

In a *dynamically typed* language, every variable name is bound only to an object. Names are bound to objects at execution time by means of assignment statements, and it is possible to bind a name to objects of different types during the execution of the program. This behaviour is not easily captured, particularly in a program logic setting, but is often used in practice and verification should aim to tackle it. A semantic analysis currently cannot alleviate this issue, as the semantics of Python is intricate and complex, and has yet to be fully defined to date.

Previous research efforts [22][23][24] have attempted to formalise a subset of Python that would be amenable to verification; however, it has been notoriously difficult and challenging to formalise and verify dynamically typed languages. Although optional static type hinting is now available for Python, "the Python runtime does not enforce function and variable type annotations. [Hints] can be used by third party tools such as type checkers, IDEs, linters, etc." [25][26]. Furthermore, it is unlikely that the ML community will constrain themselves to subsets of Python which are statically typed [27]. In addition, it is unclear how potential faults arising from dynamic languages could affect the functionality of an ML algorithm itself. No static analysis or formal verification methods exist to allow for the analysis of Python code, beyond simple linter analyses. Indeed, this is a large gap within the formal methods field that needs to be addressed immediately, given the deployment of autonomous vehicles utilizing Python.

# 4 Conclusions and recommendations

We have scoped and assessed the existing gaps and challenges of utilizing state-of-the-art static analysis and formal verification techniques to ML models that may be deployed in autonomous vehicles. We focused on directly investigating the desired behaviour (e.g., the safety property or reliability) of a system through an outcome-based approach.

Through the lens of our assurance case, we have demonstrated that research into properties such as pointwise robustness is not a priority for real time critical systems, where we are concerned with functionality, performance, reliability, operability, availability, and security. Furthermore, fragility of classification systems appears to be an intrinsic property, rather than a threat or potential vulnerability that can be patched. Additionally, these systems may be intrinsically unverifiable against the properties which are of interest to the safety of an autonomous vehicle. However, these properties can in principle be formulated for other types of ML systems present within autonomous vehicles (e.g., planning).

Finally, we have shown how existing static analysis tools can be used to build confidence against implementation errors that may propagate and affect the accuracy and functionality of the ML algorithms. However, the verification of existing ML software written in Python poses particular challenges.

In general, we believe that formal verification and static analysis are potentially important techniques for assuring classes of ML based systems, and should be considered in the assurance of any safety critical application. Below, we present recommendations that should be addressed to allow the use of formal verification and static analysis in the future assurance of ML-based safety critical systems:

1. ML-specific properties such as pointwise robustness not only fail to note real-world examples, but also how state-of-the-art verification techniques can be applied to real-time systems.
    1.1. Strategizing the use of formal methods through the lens of an assurance approach – in particular, claims, arguments, and evidence – allows one to identify the role of V&V methods and how they complement other approaches.
2. With regard to the safety assurance of an autonomous system, pointwise robustness fails to support or provide evidence for system robustness as discussed. We thus recommend:
    2.1. Creation of relevant safety specifications unique to ML algorithms, with corresponding mathematical frameworks. The noted specifications must contribute to the assurance of an AI system, specifically, the context of an assurance case (i.e., CAE).
3. Some ML algorithms (e.g., vision systems) may be intrinsically unverifiable against the properties which are of interest to the safety of an autonomous vehicle. However, other properties can in principle be formulated for other types of ML systems (e.g., planning) in autonomous vehicles.
    3.1. A collaboration between ML and verification researchers is thus needed to produce deep learning systems that are more amenable to verification, as described by [28].
    3.2. Novel formal verification techniques should address the newly defined specifications.
4. The ML lifecycle relies heavily on data processed in a complex chain of libraries and tools traditionally implemented, often in Python. It has been demonstrated that implementation in these systems may propagate and affect the accuracy and functionality of the ML algorithm itself. Formal methods can have a strong role in ensuring provenance of training and data processing. We thus recommend:

4.1. Creation of novel formal verification techniques addressing Python, and more generally, dynamically typed languages.
4.2. That organisations should consider rewriting any deployed safety critical software in a verifiable language if the appropriate analysis tools for Python are unavailable.

5. Organisation need to understand the extent to which existing integrity static analysis tools can contribute to the confidence of the development of ML algorithms. In particular, the complexities arising from choice of implementation language, e.g., issues from using C or C++, should be well understood.

# 5 Bibliography

[1] Avizienis, A., Laprie, J., Randell, B., Landwehr, C. "Basic concepts and taxonomy of dependable and secure computing," in IEEE Transactions on Dependable and Secure Computing, vol. 1, no. 1, pp. 11-33, March 2004.

[2] Papernot, N., McDaniel, P., Sinha, A., & Wellman, M. Towards the Science of Security and Privacy in Machine Learning. arXiv preprint arXiv:1611.03814, 2016.

[3] Szegedy, C., Zaremba, W., Sutskever, I., Bruna, J., Erhan, D., Goodfellow, I., & Fergus, R. Intriguing properties of neural networks. arXiv preprint arXiv:1312.6199, 2013.

[4] Goodfellow, I., Shlens, J., Szegedy, C. "Explaining and harnessing adversarial examples," in International Conference on Learning Representations. Computational and Biological Learning Society, 2015.

[5] McSherry, F. "Statistical inference considered harmful," 2016. [Online]. Available: https://github.com/frankmcsherry/blog/blob/ master/posts/2016-06-14.md last accessed March 2019.

[6] Fredrikson, M., Jha, S., Ristenpart, T. "Model inversion attacks that exploit confidence information and basic countermeasures," in Proceedings of the 22nd ACM SIGSAC Conference on Computer and Communications Security. ACM, 2015, pp. 1322–1333.

[7] Pulina, L., & Tacchella, A. An abstraction-refinement approach to verification of artificial neural networks. In International Conference on Computer Aided Verification (pp. 243-257). Springer Berlin Heidelberg, July 2010.

[8] Huang, X., Kwiatkowska, M., Wang, S., & Wu, M. Safety Verification of Deep Neural Networks. arXiv preprint arXiv:1610.06940, 2016.

[9] Katz, G., Barrett, C., Dill, D., Julian, K., & Kochenderfer, M. Reluplex: An Efficient SMT Solver for Verifying Deep Neural Networks. arXiv preprint arXiv:1702.01135, 2017.

[10] Ruan, W., Huang, X., & Kwiatkowska, M.Z. Reachability Analysis of Deep Neural Networks with Provable Guarantees. IJCAI, 2018.

[11] Schumann, J., Gupta, P., Nelson, S. On verification & validation of neural network based controllers. NASA, 2003.

[12] Can robot navigation bugs be found in simulation? An exploratory study, T Sotiropouls et al. 2017 IEEE International Conference on Software Quality, Reliability and Security (QRS).

[13] Gilmer, J., Adams, R.P., Goodfellow, I.J., Andersen, D., & Dahl, G.E. (2018). Motivating the Rules of the Game for Adversarial Example Research. CoRR, abs/1807.06732.

[14] Wang, Z. and Bovik, A. C. "Mean squared error: Love it or leave it? A new look at signal fidelity measures". In: IEEE Signal Processing Magazine 26.1 (2009), pp. 98– 117.

[15] Robustness Testing of Autonomy Software, C Hutchison et al. IEEE/ACM 40th International Conference on Software Engineering: Software Engineering in Practice Track (ICSE-SEIP), May/June 2018.

[16] "What's new in YOLO v3?" https://towardsdatascience.com/yolo-v3-object-detection-53fb7d3bfe6b last accessed March 2019.

[17] Polyspace Bug Finder, https://www.mathworks.com/products/polyspace-bug-finder.html last accessed March 2019.

[18] K Semantic Framework, Python 3.3. Accessed March, 2019. https://code.google.com/archive/p/k-python-semantics/.

[19] Shamir, A., Safran, I., Ronen, E., Dunkelman, O. A simple explanation for the existence of adversarial examples with small hamming distance. arXiv preprint arXiv:1901.10861

[20] Clark, M.B., Koutsoukos, X.D., Porter, J., Kumar, R., Pappas, G.J., Sokolsky, O., Lee, I., & Pike, L. (2013). A Study on Run Time Assurance for Complex Cyber Physical Systems.

[21] Daniel Selsam, Percy Liang, and David L. Dill. 2017. Developing bug-free machine learning systems with formal mathematics. In Proceedings of the 34th International Conference on Machine Learning - Volume 70 (ICML'17).

[22] Joe Gibbs Politz, Alejandro Martinez, Matthew Milano, Sumner Warren, Daniel Patterson, Junsong Li, Anand Chitipothu, and Shriram Krishnamurthi. 2013. Python: the full monty. SIGPLAN Not. 48, 10 (October 2013), 217-232. DOI: https://doi.org/10.1145/2544173.2509536

[23] K-Framework, Python 3.3 Semantics. Accessed November 2019. https://github.com/kframework/python-semantics.

[24] Philippa Anne Gardner, Sergio Maffeis, and Gareth David Smith. 2012. Towards a program logic for JavaScript. SIGPLAN Not. 47, 1 (January 2012), 31-44. DOI: https://doi.org/10.1145/2103621.2103663

[25] Python Docs, Typing – Support for type hints. Accessed November 2019. https://docs.python.org/3/library/typing.html.

[26] MyPy – Optional Static Typting for Python. Accessed November 2019. http://mypy-lang.org/.

[27] Tensorflow - PEP 484 Type Annotations (feature request). Accessed November 2019. https://github.com/tensorflow/tensorflow/issues/12345.

[28] Kuper, Lindsey & Katz, Guy & Gottschlich, Justin & Julian, Kyle & Barrett, Clark & Kochenderfer, Mykel. (2018). Toward Scalable Verification for Safety-Critical Deep Networks.

[29] Guy Katz, Derek A. Huang, Duligur Ibeling, Kyle Julian, Christopher Lazarus, Rachel Lim, Parth Shah, Shantanu Thakoor, Haoze Wu, Aleksandar Zeljic, David L. Dill, Mykel J. Kochenderfer, Clark W. Barrett. The Marabou Framework for Verification and Analysis of Deep Neural Networks. CAV 2019: 443-452.

[30] TIGARS Topic Note Simulation and Dynamic Testing, D5.6.5 (W3015). December 2019.

[31] Guidelines for the Use of the C Language in Critical Systems, ISBN 978-1-906400-10-1 (paperback), ISBN 978-1-906400-11-8 (PDF), March 2013. https://www.misra.org.uk/Publications/tabid/57/Default.aspx#label-comp